\documentclass[useAMS,usenatbib]{mn2e}

\usepackage{graphicx,psfig}

\usepackage{psfig}

\def\ergs{erg\,s$^{-1}$}
\def\ss{s\,s$^{-1}$}
\def\ergscm2{erg\,cm$^{-2}$s$^{-1}$}
\def\ss{s\,s$^{-1}$}

\def\1e{1E\,1547.0-5408\,}

\newcommand{\ROSAT}{{\it ROSAT}\,}
\newcommand{\Fermi}{{\it Fermi}\,}
\newcommand{\XMM}{{\it XMM--Newton}\,}

\newcommand{\CXO}{{\it Chandra}\,}

\title[Discovery of 59~ms Pulsations from 1RXS
J141256.0+792204]{Discovery
of 59~ms Pulsations from 1RXS
J141256.0+792204 (Calvera)}
\author[ ]{S. Zane$^{1}$\thanks{E-mail:
sz@mssl.ucl.ac.uk}, F. Haberl$^{2}$, G.L. Israel$^{3}$, A.
Pellizzoni$^{4}$, M. Burgay$^{4}$, R.P. Mignani$^{1}$,
\newauthor R. Turolla$^{5,1}$, A. Possenti$^{4}$,
P. Esposito$^{4,6}$, D. Champion$^{7}$, R.P. 
Eatough$^{7}$, 
\newauthor E. Barr$^{7}$, M. Kramer$^{7}$ 
\smallskip \\
$^{1}$ Mullard Space Science Laboratory, University College London,
Holmbury St. Mary, Dorking, Surrey, RH5 6NT,
UK; \\
$^2$  Max-Planck-Institut f\"ur extraterrestrische Physik, 
Giessenbachstrasse, D-85748, Garching
Germany
\\
$^3$  INFN-Osservatorio Astronomico di Roma, via Frascati 33, I-00040 
Monteporzio Catone, Italy
 \\
$^4$  INAF-Osservatorio Astronomico di Cagliari, localit\`a Poggio dei 
Pini, strada
54, I-09012 Capoterra, Italy
\\
$^5$ Department of Physics, University of Padova, via
Marzolo 8, 35131 Padova, Italy \\
$^6$ 
INFN - Istituto Nazionale di Fisica Nucleare, sezione di Pavia, via
A.~Bassi 6, I-27100 Pavia, Italy
\\ 
$^7$ Max-Planck-Institut  f\"ur Radioastronomie, Auf dem H\"ugel 69, 
53121 Bonn, Germany}

\begin{document}

\date{}

\pagerange{\pageref{firstpage}--\pageref{lastpage}} \pubyear{2007}

\maketitle

\label{firstpage}

\begin{abstract}
We report on the results of a multi-wavelength study of the 
compact object candidate 1RXS J141256.0+792204 (Calvera). Calvera was 
observed in the
X-rays with \XMM/EPIC twice for a total exposure time of $\sim 50$~ks. The 
source 
spectrum is thermal and well reproduced by a two component
model composed of either two (absorbed) hydrogen atmosphere models, or two
blackbodies with temperatures 
$kT_1\sim 55/150$~eV, $kT_2\sim 80/250$~eV, respectively (as measured at 
infinity). Evidence was 
found for an 
absorption
feature at $\sim 0.65$ keV while no power-law high-energy tail is 
statistically
required. Using pn and MOS data we
discovered pulsations in the X-ray emission at a period $P=59.2$ ms. The
detection is highly significant ($\ga 11\sigma$), and 
unambiguously confirms  the 
neutron star nature of Calvera. The pulse profile is
nearly sinusoidal, with a pulsed fraction of $\sim 18\%$. 
We looked for the timing
signature of Calvera in the \Fermi\ Large Area Telescope (LAT) 
database and found a 
significant
($\sim 5 \sigma$) pulsed signal at a period coincident with the X-ray 
value.
The gamma-ray timing analysis yielded a tight upper limit on the period
derivative, $\dot P < 5\times 10^{-18}$~\ss ($\dot E_{rot}  
<10^{33}$~\ergs, $B<5\times 10^{10}$~G for magneto- dipolar
spin-down). Radio searches at 1.36~GHz with the
100-m Effelsberg radio telescope yielded negative results, with a deep 
upper 
limit on
the pulsed flux of $0.05$ mJy. Diffuse, soft ($< 1$ keV) X-ray emission
about 13' west of the Calvera position is present both in our 
pointed observations and in archive \ROSAT\ all-sky  survey images, but 
is unlikely associated with the X-ray pulsar. Its spectrum 
is
compatible with an old supernova remnant (SNR); no evidence for diffuse 
emission in the radio and optical bands was found. 
The most likely interpretations are that Calvera is either a central 
compact object escaped from a SNR or a mildly recycled pulsar; in 
both cases the source would be the first ever member of the class detected 
at gamma-ray energies.
\end{abstract}


\section{Introduction}
\label{intro}

Isolated neutron stars (NSs)  have been for a long time associated with
radio pulsars. It was only during the last two decades, thanks to
multi-wavelength observations from both space and ground-based
instruments, that our picture of the Galactic NS population changed.
These observations unveiled the existence of different types of
isolated NSs which
are radio quiet or have radio properties quite at variance with those of
ordinary pulsars. Examples are the soft gamma repeaters and the
anomalous X-ray pulsars \cite[the likely magnetars;
e.g.][]{wt06,mere08}, the central compact objects \cite[CCOs;
e.g.][]{pav04,deluca08} in supernova 
remnants (SNR),
the
seven dim \ROSAT\ sources \cite[X-ray Dim Isolated NSs, XDINSs; 
e.g.][]{h07,tur09} 
and the
more
recently discovered rotating radio transients \cite[RRATs;][]{mcl06}.
Observations at X-ray, gamma-ray and optical wavelengths have been
crucial in revealing the rich variety of NS properties and their
phenomenology.

Only a limited number ($\le 20$, in comparison with $\sim$2000 known 
ordinary radio pulsars) of sources in each of these new isolated NS 
classes has 
been
identified so far.  This makes establishing links between
different groups hazardous and no unitary picture of the Galactic NSs
has emerged to date.  The search for new members of each
class is therefore of the utmost importance, as is the study of
objects with properties intermediate between those of known 
groups, which 
could 
provide
the much needed missing links.

In this respect, it appears very promising that two new,
apparently radio quiet X-ray sources have been
recently proposed as isolated NS candidates. These are
1RXS J141256.0+792204  \cite[dubbed Calvera;][]{he07,rut08,sh09}, which
was found in the \ROSAT\ Bright Source Catalogue
\cite[RBSC,][]{vog99}, and, at fainter flux levels, 2XMM~J104608.7-594306,
discovered with \XMM\ \citep{pir09}.

Calvera, in particular, is a puzzling source and its 
interpretation represented a  conundrum (see
\S~\ref{storia}). At variance with most known radio quiet isolated
NSs, the source is at high Galactic latitude, and it exhibits a quite hot,
thermal
spectrum with little absorption. Calvera was first selected in the
RBSC and identified as a possible 
isolated
NS candidate on the basis of its large X-ray-to-optical flux ratio.  
The source was then observed with the {\it Swift}-X-Ray Telescope (XRT) 
and with \CXO\ 
(see \S~\ref{storia} for details of the past observations, and
references therein). In particular, the  
first, short  ($\sim$2 ks) \CXO\ pointing (obs. ID: 8508) provided a 
refined X-ray 
position and
follow-up Gemini observations confirmed the lack of optical counterparts 
down to $g \sim 26.3$,
implying $F_X/F_{opt} > 9000$ \citep{rut08}. However, despite the
improved spectral information, these data were insufficient to
discriminate among the different interpretations for its nature.

Clearly, better spectral and timing information is crucial to
unambiguously classify the source.
Here, we present a multi-wavelength study of Calvera, based on two new
\XMM\ observations, on the analysis of publicly available \Fermi-LAT
data, and on a new radio observation taken at the 100-m Effelsberg 
telescope. We also used \CXO\ and \ROSAT\ archival data. 
The paper is organised
as follows. We first summarise the results from the past 
observations
and the proposed scenarios in \S~\ref{storia}. Spectral 
and timing
results from the
 \XMM \ and \Fermi-LAT
observations are presented in
\S~\ref{data}, while
the  radio observations are summarised in \S~\ref{radio}.
A search for
potential sources of diffuse X-ray or radio emission in the proximity of
Calvera is presented in \S~\ref{diff}, discussion and conclusions 
follow in
\S~\ref{disc},\ref{conc}.

\section{Past observations and proposed scenarios}
\label{storia}

The first X-ray observations of Calvera, taken with the {\it Swift}-XRT, 
had  very limited  counting statistics  (74  source counts  in  the 
0.3--10  keV
range) and only allowed a very preliminary spectral analysis
\citep{rut08}:
the  spectrum appeared  rather hard, but the spectral  model was
virtually  unconstrained.  A  blackbody (BB) fit  gave temperature 
$kT_{BB}  \sim 215$ eV,  radius $R_{BB}\sim 7$  km (for  a 10  kpc 
distance),
while a power-law (PL) model  yielded a photon index  $\Gamma\sim 
2.8$. The 0.3-10 keV flux 
was
$1.2\times 10^{-12}$  and $2.5\times 10^{-13}$~\ergscm2 for
the  two   models,  respectively.  A  thermal   bremsstrahlung  and  a
Raymond-Smith plasma model  were also tried, providing  temperatures
of
$\sim  0.8$ and  $\sim  1.5$  keV.  No  estimate  of the  interstellar
absorption  was  derived (in all fits $N_H$ was held fixed to the 
total Galactic value obtained from  \cite{kal05}), so the 
source distance was unconstrained.

On the basis of the very limited spectral information obtained with the
{\it Swift}\ data,
\cite{rut08} discussed several intriguing possibilities for the nature of
Calvera. However, they also concluded that none of them was without
problems.  For instance, they
proposed that the source may be an XDINS, but hotter
than the other members of the class ($kT_{BB} \sim 40$--100 eV). In
this case,
under
the arbitrary assumption that the emitting radius is the same for
all objects of the class, the observed flux would imply a distance of 
$\sim 2-8$ kpc
from
the Sun and a height of $\sim 1.3-5.1$~kpc above the Galactic plane.
Taken at face value, this appears hardly compatible with the star's age as 
derived
from its temperature (assuming standard cooling models), unless either
Calvera has an unprecedented high velocity (a few thousand km\,s$^{-1}$), 
or 
did not cool according to conventional models, or it has been re-heated by
some unknown mechanisms.  

A magnetar identification was also recognised to
be problematic.
For an average quiescent magnetar luminosity of $\approx 10^{35}$~\ergs, 
the flux  inferred from the  best fit of  the {\it
Swift}-XRT spectrum would imply  a distance of at least 66~kpc, 
and  a height 
of
40 kpc above the Galactic plane.  Calvera would then be the first magnetar
to belong to a halo
population. Although evidence of massive compact object
formation in the halo has been found \cite[e.g.][]{mira01}, this
interpretation was considered statistically unlikely. On the other
hand, a Galactic
plane population
origin would require a luminosity fainter by about 5 orders of magnitude,
in which case Calvera would be a candidate for the faintest, and maybe the
oldest, identified magnetar.

\cite{rut08} also considered the idea  that Calvera might be a CCO.
In this case, assuming an average NS 
velocity of $\approx$ 400 km s$^{-1}$ and a cooling age of 0.5 Myrs
\citep{page04}, the present location of Calvera
would imply a distance of $\sim
300$~pc which would make it a factor $\approx 10$ fainter than other known
CCOs. 

The last alternative proposed by \cite{rut08}, was an old
millisecond pulsar. 
This scenario would place the source at a distance of a few hundreds pc, 
the exact value depending
on the assumed size of the hot polar caps. Interestingly, for such a small
distance, the derived upper limit in the optical of $g> 26.3$ reported by 
\cite{rut08} is deep enough to rule out both a white or a red dwarf
companion. If this is the case, Calvera would thus be one of the few
solitary field millisecond pulsars discovered so far (about 18  
known isolated millisecond pulsars in the Galactic plane, comprising
roughly 30\% of the population).

In order to shed further light on these possibilities,
\cite{sh09}
re-observed the source with \CXO\ for 30~ks on 2008 April 8 (obs. ID: 
9141). This longer
observation allowed them to exclude (absorbed) single-component spectral
models, as
a BB, a PL or a pure hydrogen atmospheric model
\cite[NSA;][]{zav96}, since they all gave
unacceptably large values of $\chi^2$ and, in some cases, a value of the
column
density larger than the Galactic one in the source direction
\cite[$N_{gal} = 2.65 \times 10^{20}$~cm$^2$;][]{kal05}. Instead, 
these authors
proposed as best fit a NSA model (with effective temperature of $\approx
110-120$~eV) combined with a spectral feature, either an edge or more 
likely an emission line, at $\sim 0.5-0.6$~keV. 
The source did not show signs of variability neither on short
nor long ($> 1$~year) timescales, and pulsations have not been detected
down to a 3$\sigma$ limit of 20-30\% in pulsed fraction (defined as the
semi-amplitude of the sinusoidal modulation divided by the mean source
count rate) in the period range  between 0.9\,s and 10$^4$\,s.
The absence of a non-thermal
spectral component and the relatively low limit on the pulsed fraction for
periods of a few seconds made a magnetar interpretation unlikely, but the
other possibilities still remained open. 
Therefore, despite the better spectrum, the \CXO\ data (obs. ID: 
9141) did not provide sufficient clues to solve the conundrum 
proposed by \cite{rut08} about the nature of the source.

\section{X-ray and Gamma-ray Observations}
\label{data}

\subsection{Source position and association}
\label{pos}

\begin{figure*}
\vbox{ \hbox{
\psfig{figure=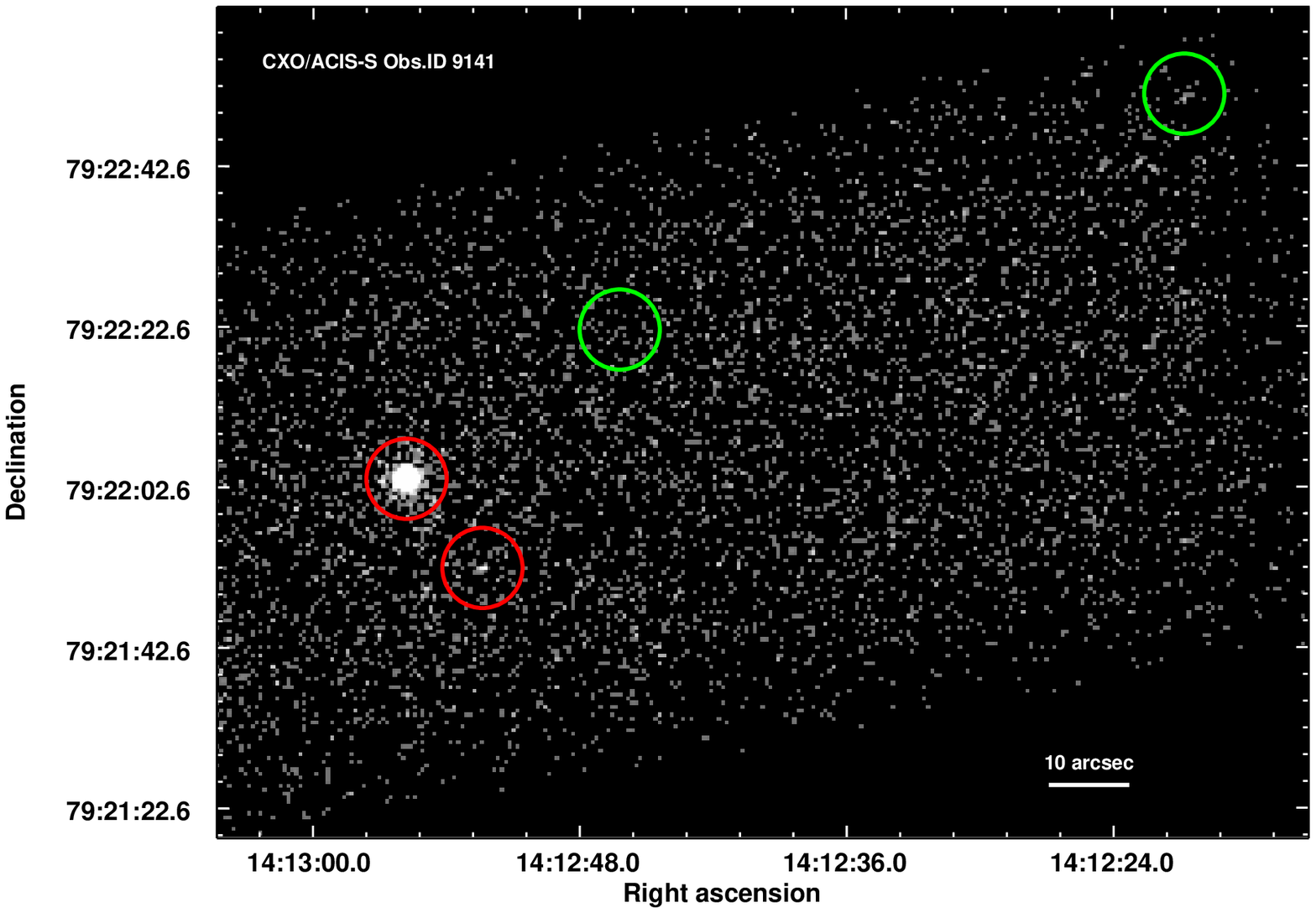,width=0.52\hsize}
\hspace{0.02\hsize}
\psfig{figure=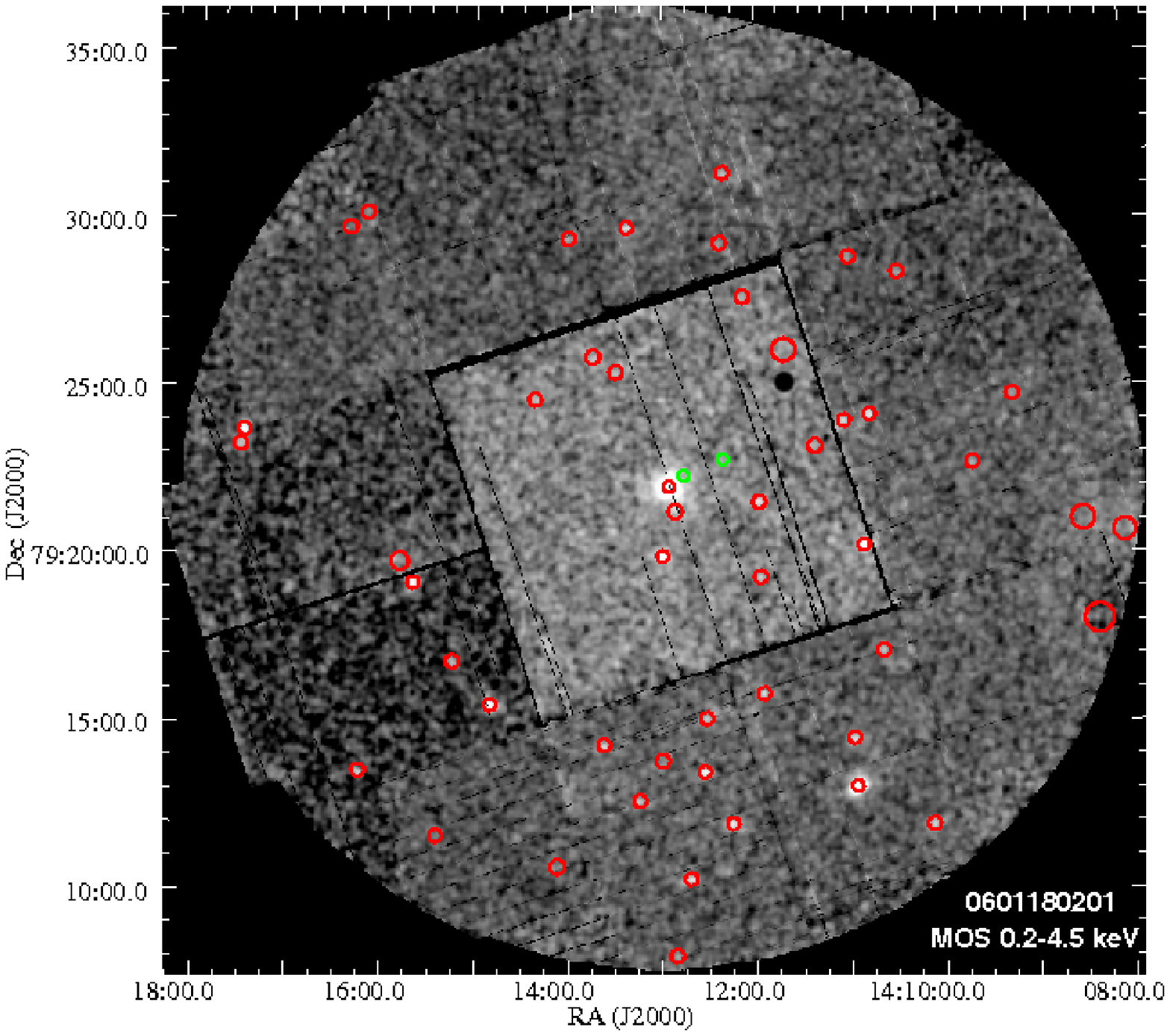,width=0.48\hsize}}
\caption
{{\it Left panel:} \CXO\ image of the field taken in 2008 
(obs. ID: 9141). Red circles
mark the two 
detected sources: Calvera ($\alpha =  
14^{\rm h} 12^{\rm m} 55\fs83$, $\delta=  
79^\circ  22\arcmin
3\farcs$7, uncertainty = 0.63\arcsec at 90\% confidence level, cts/s = 
0.180 $\pm$ 
0.003) 
and  
CXOU\, J141252.41+792152.60 (
$\alpha = 14^{\rm   h} 12^{\rm  m} 52\fs41$, $\delta =  79^\circ 
21^{\rm m}  
52\farcs6$, 
uncertainty = 1.41\arcsec  
at 
90\% confidence 
level, cts/s =
0.00052 $ \pm$ 0.00018).  Note
that these two sources are not resolved in the \XMM\ data, but, given the
low count rate of the second one, we do not expect a significant
contamination in our spectral and timing analysis.
The green circles mark the positions of the two sources used by 
Shevchuk et al.~(2009) for the bore sight correction; note they are 
undetected in
our analysis of the \CXO\ data.
{\it Right panel:} Combined EPIC MOS1/2 image  in the 0.2-4.5 keV
band,  with the detected sources marked by red circles.
MOS1 was in imaging mode, covering the target
and the surrounding field (central CCD + 5 outer CCDs), while MOS2 was in
timing mode (imaging for outer 6 CCDs). Green circles
correspond to the green circles on the \CXO\ image.
\label{figfield}}}
\end{figure*}

We started our analysis by 
first accurately re-assessing the source position determination. 
This is  particularly important because a large positional 
error precludes firmly ruling out the association with 
the close-by ($\sim 0.9$\arcsec) $g\sim 24.8$ Star A \citep{rut08}.

Based on a \CXO/HRC-I observation (obs. ID: 8508), \cite{rut08} 
gave:  
$\alpha=14^{\rm h} 
12^{\rm m} 55\fs885$, $\delta=+79^\circ 22\arcmin 04\farcs10$, with a 90\% 
uncertainty of $0\farcs57$, after applying a bore sight correction of 
$\Delta \alpha = -0\farcs04 \pm 0\farcs22$ and $\Delta \delta =+0\farcs95 
\pm 0\farcs22$ to the \CXO\ astrometry.\footnote{As the authors noted, 
this large correction was probably due to a systematic 0\farcs4 offset in 
the absolute \CXO\ astrometry affecting data taken after 2006 December 
1.} To this aim, they used as a reference the optical coordinates of a 
second source, CXOU\, J141259.43+791958, detected in the HRC-I field, 
derived from those of its putative counterpart identified in their Gemini 
images, calibrated with the USNO-B1.0 catalogue \citep{monet03}. However, 
this procedure comes with some caveats.

To verify the value of the \cite{rut08} coordinates, we retrieved the 
HRC-I data set (obs. ID: 8508) from the public \CXO\ archive. First of 
all, 
we found that their reference X-ray source has a count rate of $\approx$ 
0.005~cts/s ($\approx 2.5 \sigma$ detection significance), which lowers 
the 
accuracy of the bore sight correction.  Furthermore, using only one 
reference source can introduce systematics due to a proper motion of its 
putative optical counterpart.  Since the default reference epoch for the 
USNO-B1.0 coordinates is 2000.0\footnote{USNO-B1.0 coordinates are 
extrapolated to epoch 2000.0 from the mean epoch of observations according 
to the proper motions computed by the USNO-B1.0 pipeline.  However, the 
uncertainty due to the proper motion extrapolations is not accounted for 
in the nominal coordinate accuracy.  This implies larger errors on any 
USNO-B1.0-based astrometric solution and on the coordinates thereby 
determined.} while the epoch of the \CXO/HRC-I observation (obs. ID: 
8508) is 2007.13, the 
effect on the bore sight correction can be significant.  We identified the 
putative counterpart of the reference X-ray source with the USNO-B1.0 star 
1693-0051235 ($B=20.51; R=19.4$).  According to the catalogue, this star 
has a proper motion $\mu_{\alpha} = -6 \pm 4$ mas yr$^{-1}$ and 
$\mu_{\delta} = 6 \pm 1$ mas yr$^{-1}$, with $\pm 4$ mas yr$^{-1}$ being a 
more realistic uncertainty for the latter value \cite[see  
e.g.][]{Gould03}.  This would imply an offset of $\Delta \alpha \sim +42 
\pm 28$ mas and $\Delta \delta = -42 \pm 28$ mas in the bore sight 
correction, apparently not accounted for in \cite{rut08}.  While this 
offset would be negligible within the error budget of the \CXO\ and of the 
USNO-B1.0-based astrometry calibration of the Gemini images \citep{rut08}, 
we can not rule out that the star has a larger proper motion. We queried 
other astrometric catalogues to verify the USNO-B1.0 proper motion. 
Unfortunately, the star is too faint to be in the UCAC-3 catalogue 
\citep{zac10} while it appears in the PPMXL \citep{roeser10} which gives 
$\mu_{\alpha} = -7.1 \pm 5.6$ mas yr$^{-1}$ and $\mu_{\delta} = 6.1 \pm 
5.6$ mas yr$^{-1}$, consistent with the USNO-B1.0 value although not 
statistically significant yet.  From this value, we set a $3 \sigma$ upper 
limit on the star proper motion of $\mu_{\alpha} = 24 $ mas yr$^{-1}$ and 
$\mu_{\delta} = 23$ mas yr$^{-1}$.  Thus, we estimate an additional 
uncertainty of $\sim 0\farcs17$ (per coordinate) to the one-star bore 
sight correction of \cite{rut08}, rising its total uncertainty to 
$0\farcs28$ (per coordinate).

We independently measured the coordinates of Calvera in the HRC-I dataset, 
after it was corrected for the reported offset in the absolute \CXO\ 
astrometry.  Our best fit coordinates are then $\alpha=14^{\rm h} 12^{\rm 
m} 55\fs64$, $\delta=+79^\circ 22\arcmin 03\farcs7$, with a 90\% 
confidence error of 0\farcs6. Given the re-assessed uncertainty on the 
one-star bore sight correction (see above), we decided not to apply it to 
the measured coordinates of Calvera.

A second \CXO\ observation of Calvera was performed with the ACIS-S 
detector \citep[][obs. ID: 9141]{sh09}.  We retrieved the ACIS-S dataset 
from the public \CXO\ archive and we measured the position of Calvera as 
$\alpha=14^{\rm h} 12^{\rm m} 55\fs84$, $\delta=+79^\circ 22\arcmin 
03\farcs7$, with a 90\% confidence error of 0\farcs6. This position is 
consistent with that measured by us using the HRC-I data set and, 
obviously, virtually identical to that obtained by \cite{sh09} from the uncorrected 
\CXO\ astrometry. We noticed that \cite{sh09} applied the bore sight 
correction of the ACIS-S image, using the position of the optical 
counterparts to two field sources, CXOU\, J141220.78+792251.6 and CXOU\, 
J141246.23+792222.3. They then obtained $\alpha=14^{\rm h} 12^{\rm m} 
55\fs76$, $\delta=+79^\circ 22\arcmin 03\farcs4$ with a quoted 90\% 
uncertainty of 0\farcs31 and 0\farcs3 in right ascension and declination, 
respectively. However, we could not detect these two sources in the ACIS-S 
(above a detection threshold of 3$\sigma$, see Fig.~\ref{figfield}, left 
panel).  We confirm, instead, the marginal detection of a faint source, 
CXOU\, J141252.41+792152.60, in the ACIS-S data located $\sim 20\arcsec$ 
south west of Calvera, which was not detected in the HRC-I data (obs. ID: 
8508), while the 
HRC-I source CXOU\, J141259.43+791958 falls out of the ACIS-S field of 
view.  Because of the non-detection of the two sources claimed by 
\cite{sh09} in the ACIS-S data, we used the nominal \CXO\ position in the 
present analysis.

\begin{table}
\caption[]{\XMM\ EPIC observations of Calvera.}
\begin{tabular}{llrl}
\hline\noalign{\smallskip}
\multicolumn{1}{c}{Observation} &
\multicolumn{1}{c}{EPIC$^{(a)}$} &
\multicolumn{1}{c}{Start -- End time (UT)} &
\multicolumn{1}{c}{Exp.$^{(b)}$} \\

\multicolumn{1}{c}{ID} &
\multicolumn{1}{c}{} &
\multicolumn{1}{c}{} &
\multicolumn{1}{c}{(ks)} \\

\noalign{\smallskip}\hline\noalign{\smallskip}
 0601180101 & pn SW & 2009-08-31 07:14--15:09 & 14.0/13.94 \\
            & M1 FF &        07:08--15:09 & 19.6/19.65 \\
            & M2 TU &        07:08--15:05 & 20.0/ --   \\
 0601180201 & pn SW & 2009-10-10 04:15--12:27 & 14.5/19.48 \\
            & M1 FF &        04:09--12:26 & 20.3/27.44 \\
            & M2 TU &        04:09--12:22 & 20.4/ --   \\
\noalign{\smallskip}\hline\noalign{\smallskip}
\end{tabular}

$^{(a)}$ Instrument configurations: pn SW: Small Window CCD readout mode
with 6~ms frame time; MOS1 (M1) FF: Full Frame, 2.6~s;
         MOS2 (M2)  TU: Timing Uncompressed, 1.75~ms; The thin optical 
blocking filter was used for all cameras.\\
$^{(b)}$ Net exposures used for timing and spectral analysis (dead-time 
corrected).
\label{tabobs}
\end{table}

Finally, for the sake of completeness we checked our \XMM\ observations.  
\XMM\ observed Calvera twice in 2009, on August 31 and October 10, for a 
total exposure of about 47 ks. The observations were performed with the 
EPIC pn \citep{2001A&A...365L..18S} and MOS \citep{2001A&A...365L..27T} 
cameras in different CCD read out modes (see Table~\ref{tabobs}).  The net 
exposure from the first observation (obs. ID: 0601180101) 
was short: less sources were detected 
and no bore sight correction was performed. From the second observation 
(obs. ID: 0601180201), 
we obtained $\alpha=14^{\rm h} 12^{\rm m} 55\fs48$, $\delta=+79^\circ 
22\arcmin 03\farcs3$, as computed by the EPIC pipeline (1$\sigma$ 
statistical error of 0\farcs21)  which computes the bore sight correction 
from the positions of X-ray sources in the field matched against those of 
optical sources in, e.g. the USNO-B1.0, NED, and NOMAD catalogues. In this 
case, the pipeline finds 35 X-ray sources with optical matches. However, 
we visually inspected the X-ray source position on the DSS-2 images and we 
found possible matches with 14 optical sources, of which only 4 are listed 
both in the GSC-2 \citep{las08} and in the 2MASS \citep{sk06} catalogues.  
By matching the pixel-to-sky coordinates of the X-ray sources and their 
associated 2MASS/GSC-2 counterparts we derived an astrometric solution 
with an rms of 0\farcs88 and we determined $\alpha=14^{\rm h} 12^{\rm m} 
55\fs68$, $\delta=+79^\circ 22\arcmin 04\farcs2$ for Calvera, formally 
more consistent with the coordinates obtained from the \CXO\ data sets.

We notice that the two sources used by \cite{sh09} for the \CXO\ bore 
sight are also not detected by \XMM\ (see Fig.~\ref{figfield}, right 
panel).  From the sensitivity map of observation 0601180201 (pipeline), 
the $3 \sigma$ upper limit for the two sources is 0.005~cts/s (in the EPIC 
total band 0.2-12~keV), a factor of 30 lower than the count rate of 
Calvera. This converts to a flux of $\approx 3 \times 10^{-14}$~\ergscm2, 
with sizeable uncertainties depending on the assumed spectral model.

In conclusion, through our astrometry analysis we conclude that the most 
reliable coordinates of Calvera are those derived from the \CXO\ 
astrometry 
($\alpha=14^{\rm h} 12^{\rm m} 55\fs84$, $\delta=+79^\circ 22\arcmin
03\farcs7$, with a 90\% confidence error of 0\farcs6) 
i.e.  without applying uncertain bore sight corrections, and 
they indeed confirm that Calvera is not associated with star A from 
\cite{rut08}. 
Moreover, we note that the uncertainty on the Calvera
coordinates is small enough not to hamper a sensitive search for coherent
signals in our timing analyses (see \S~\ref{datafermi}).

\subsection{XMM-Newton data analysis}
\label{xmmdata}

\begin{table*}
\caption{Summary of spectral fits.}
\begin{tabular}{lclccllccl}
\hline
Model & $\chi_r^2$ / d.o.f.& $N_H$& $\Gamma$ & $kT^{(a)}$ &
$E_{edge}/E_{line}$ & $\tau/\sigma$ & F$_{obs}^{(b)}$ & F$_{bol}^{(c)}$ & Comment \\
& &  10$^{20}$  & &   &    &     & 10$^{-13}$ & 10$^{-12}$ & \\
& &  cm$^{-2}$  & & eV& keV&
& \ergscm2 & \ergscm2 &
\\
\hline
PL$^{(d)}$          & 1.843/326 & 11.9         & 3.4     &                     
&             &           & 9.12 &  --       & $N_H>N_{gal}^{(e)}$ \\
PL + BB             & 1.055/324 &  8.9$\pm1.0$ & 3.8$\pm0.3$ & 227$\pm8$               
&             &           & 8.58 & 0.51      &  $N_H>N_{gal}$ \\
PL + NSA            & 1.051/324 &  9.1$\pm0.8$ & 4.3$\pm0.5$ & 111$\pm7$               
&             &           & 8.49 & 1.28      & 
 $N_H>N_{gal}$ \\
\noalign{\smallskip}\noalign{\smallskip}
BB$^{(d)}$          & 2.258/326 &  0.0         &         & 198                 
&             &           & 8.12 & 0.84      &        \\
2BB                 & 1.053/324 &  4.6$\pm1.0$ &         & 
93$\pm8$/242$\pm9$          &             &           & 8.40 & 0.79/0.67 &  
$N_H>N_{gal}$ \\
2BB                 & 1.096/325 &  2.7$^{(f)}$ &         & 
110$\pm6$/253$\pm10$        &             &           & 8.48 & 0.58/0.59 &            \\
BB $\times$ edge    & 1.486/324 &  0.5$\pm0.3$ &         & 201$\pm3$               
& 0.66$^{+12}_{-17}$  & 0.67$\pm0.08$ & 8.30 & 1.02      &        \\
BB -- gauss         & 1.631/324 &  0.5$\pm0.3$ &         & 196$\pm3$               
& 0.76$\pm0.02$       & 0.1$^{(f)}$   & 8.23 & 0.97      & EW$= -83$~eV\\
2BB $\times$ edge   & 1.051/323 &  2.7$^{(f)}$ &         & 
123$\pm11$/254$\pm15$       & 0.63$\pm0.03$       & 0.26$\pm0.11$ & 8.47 & 0.61/0.58 &            \\
2BB -- gauss        & 1.015/322 &  2.7$^{(f)}$ &             & 
140$\pm14$/283$\pm30$           & 0.66$\pm0.05$       & 0.13$\pm0.05$ & 
8.51 & 0.82/0.41 & EW$ = -79$~eV\\
\noalign{\smallskip}\noalign{\smallskip}
NSA                 & 1.347/326 &  1.8$\pm0.4$ &         & 98$\pm3$                
&             &           & 8.37 & 1.65      &        \\
2NSA                & 1.042/324 &  6.7$\pm1.3$ &         & 
$27^{+6}_{-2}/118^{+7}_{-10}$   &             &           & 8.47 & 
1.38/3.00 &  $N_H>N_{gal}$ \\
2NSA                & 1.156/325 &  2.7$^{(f)}$ &             & 
$67^{+7}_{-12}/150^{+12}_{-20}$ &                     &               & 8.56 & 0.76/1.16 &        \\
NSA $\times$ edge   & 1.079/324 &  2.6$\pm0.5$ &         & 97$\pm3$                
& 0.65$\pm0.02$       & 0.40$\pm0.08$ & 8.43 & 1.92      &        \\
NSA + gauss         & 1.304/325 &  1.5$\pm0.4$ &         & 101$^{+4}_{-2}$             
& 0.53$^{(f)}$        & 0.0$^{(f)}$   & 8.40 & 1.59      & EW$ = 12$~eV \\
2NSA $\times$ edge  & 1.053/323 &  2.7$^{(f)}$ &             & 
82$^{+15}_{-17}$/154$^{+200}_{-40}$ & 0.64$\pm0.03$   & 0.31$\pm0.09$ & 8.53 & 0.46/1.45 &        \\
\hline
\end{tabular}
$^{(a)}$ BB and NSA temperatures are evaluated at infinity and at 
the NS surface, respectively. It is $T_\infty = T/(1 + z)$, see text for 
details. 
$^{(b)}$ Observed flux in the 0.2--10 keV band derived from the EPIC pn spectrum of the first observation.
$^{(c)}$ Bolometric flux for the thermal components (NSA and BB) derived 
from the EPIC pn spectrum of observation 0601180101.
$^{(d)}$ No error estimate due to bad fit. 
$^{(e)}$ $N_{gal} = 2.65 \times 
10^{20}$~cm$^2$, \cite{kal05}.  
$^{(f)}$ Parameter fixed in the fit.
\label{tabspec}
\end{table*}

For both, spectral and timing analysis we used the latest \XMM\
Science Analysis Software (SAS), version 10.0.1. 
We checked that the pn event files 
produced by ``epchain'' were clean of
unrecognised time jumps before we applied an event time randomisation.

We extracted source spectra for Calvera from the imaging mode data of 
pn and MOS1
using
a circular region around the target position
with 15\arcsec\ radius and background spectra from a nearby source-free region
of same size. Based on its low number of counts (see
Fig.~\ref{figfield} and caption),  
CXOU\, J141252.41+792152.60 is not expected to contaminate Calvera's
data.
Since the spectral response in MOS timing mode is currently not well
calibrated, we did not use the MOS2 timing data for spectral analysis.
For spectra we selected single pixel events (corresponding
to PATTERN=0) for pn and all valid events for MOS (PATTERN=0-12),
excluding bad
CCD pixels and columns (FLAG=0). We used XSPEC version 12.5.0 for spectral
modelling.

The same pn and MOS event lists considered for the spectral analysis were
also used as input for the timing analysis (with different background
screening criteria, see Table~\ref{tabobs}).
In addition, in order to obtain better temporal resolution, we used also
the MOS2 data. MOS2 was operated in timing mode; in 
this mode data from
the
central CCD are collapsed into a one-dimensional row to achieve a $\sim$1.75\,ms
time resolution. Therefore, instead of 
a circular extraction region, we used a rectangular box of 20 pixel width.  The local background was measured in a
region of the images far from the extracted source itself and taking care 
that no other source falls within the extraction regions by chance 
(as we already noticed, the  \CXO\  source CXOU\, J141252.41+792152.60 is 
unresolved 
from Calvera, see \S~\ref{pos}). 
The arrival
times of the photons were corrected to the barycenter  of the Solar
System using the  JPL DE405 Solar System ephemeris \citep{de405} and the 
SAS task \textsc{``barycen''}.

\subsection{XMM-Newton spectral analysis: pulse averaged spectra}
\label{spec}

\begin{figure*}
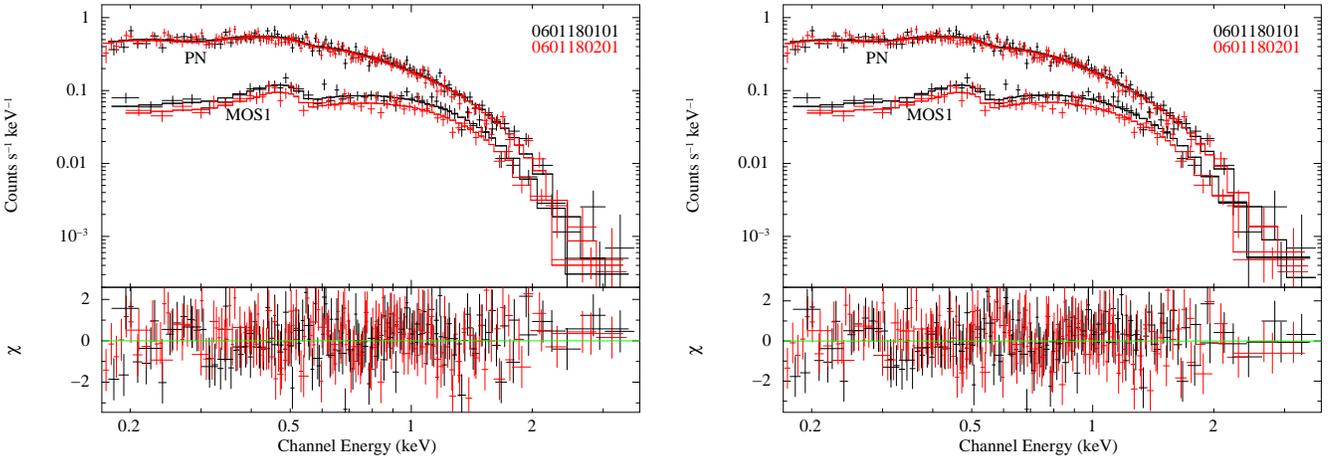

\vbox{ \hbox{
\psfig{figure=fig3.ps,width=0.48\hsize,angle=270}
\hspace{0.02\hsize}
\psfig{figure=fig4.ps,width=0.48\hsize,angle=270}}
\caption
{{\it Left panel:} EPIC pn and MOS1 spectra fitted with a double
blackbody model. 
{\it Right panel:} EPIC pn and MOS1 spectra modelled with a double
NSA model. Both figures show best-fitting models in which the
value of $N_H$ has been left free.
See Table~\ref{tabspec} and text for details. \label{fig1}}}
\end{figure*}

Spectra were accumulated from the pn and MOS1 event lists after removing
those
time intervals affected by solar proton flares resulting in an effective 
exposure time
of $\sim$14\,ks (pn) and $\sim$19.7\,ks (MOS1; first observation, 
obs. ID: 0601180101),
19.5\,ks (pn) and
27.5\,ks (MOS1; second observation, obs. ID: 0601180201).

We started the analysis considering an absorbed double BB model
and fitting the
two pn and the two MOS1 spectra
simultaneously,
allowing only a renormalisation factor to account for cross-calibration
uncertainties between the detectors and possible time-variability of the
source. To account for interstellar absorption, here and in the following
we adopted the elemental abundances of \citet[][model ``phabs'' in
XSPEC]{2000ApJ...542..914W}. We note here that the used  metal 
abundances 
influence the derived equivalent
hydrogen column density and can introduce systematic uncertainties 
when comparing its value with H\textsc{I}-derived column densities.
We did not find flux variations greater than 3\% between the two
observations, which is consistent, considering the statistical
uncertainties, with a steady source flux. We then
performed an independent fit of the spectra from
the first and the second
observation (still joining pn and MOS data within each epoch). The resulting
best-fit parameters, from the two epochs, were consistent
within their statistical errors.

Since the source showed neither significant flux 
variations nor spectral changes
between the two \XMM\ observations, in the rest of the analysis we
fitted all pn and MOS data together in order to obtain the best signal to
noise ratio. We tested several single, double and triple
component spectral models, based on different combinations of 
BBs, PLs, NS atmosphere models (NSA) and absorption edges or
Gaussian lines. In the NSA model, we used $R_{star}=12$~km and
$M_{star} = 1.4 M_{sun}$ as input parameters for the NS radius and mass.
The results are summarised in Table~\ref{tabspec}, where we report only 
the fluxes
inferred from the pn data of observation  0601180101. As mentioned 
above, 
we do not see significant flux variations between the two 
observations: computing an average pn flux for the two observations,  
weighted with the exposures, gives a value 0.98 times smaller than that 
reported in Table~\ref{tabspec}. Also, the
fluxes obtained from the MOS1 spectra are consistent with those listed in the
table. We warn the reader that the parameters reported in  
Table~\ref{tabspec} have a different physical meaning inasmuch the best 
fit temperatures obtained by the BB fit are 
measured at infinity, while those from the NSA model are measured at the 
neutron star starface. The two quantities are related by $T_\infty = 
T/(1+z)$ 
where  $ 1+z \equiv 1/\sqrt{ 1- 2.952M_{star}/R_{star}}$ is the 
gravitational redshift factor.   

In agreement with \cite{sh09}, we find that single component models do not
provide a good representation of the Calvera spectrum. However, the advantage
of the EPIC spectra is the higher efficiency of the instruments below 500~eV,
where the  \CXO\  spectrum has very few counts. Fitting the  \XMM\
data with one component models leaves  a low energy excess,  that can be
accounted for by a second emission component.
Two-component models including a PL (as in the combinations PL+BB, PL+NSA)
demonstrate a problem:  the additional PL is always
very steep and represents the soft part of the spectrum (requiring a high
interstellar absorption), instead of a hard tail as
usually seen from isolated NSs with a non-thermal component 
\citep{deluca05}. Instead, we 
found
that the best two-component fits are composed of either, two thermal
components (BB+BB, NSA+NSA), or an NSA+edge model.
One difference is that,  while the NSA+edge fit
gives a value of $N_H$ consistent with the Galactic one \cite[as in the
case of the  \CXO\  data,][]{sh09}, the fits with two thermal
components give a larger value, which may suggest the presence of a local
absorption component.  However, fixing the $N_H$ at the Galactic value
does
not deteriorate the fits considerably (see, again, Table~\ref{tabspec}).
Data and best fitting models based on two thermal components (BB+BB and NSA+NSA
with free $N_H$) are shown in Fig.~\ref{fig1}.
Unfortunately,
based on present spectral data only, it is not possible to
discriminate between  a picture
in which the spectrum originates from two zones of the NS surface (with
temperatures at infinity of
$\sim 80$ and 250~eV if the emission is modelled with blackbodies, or
$\sim 55$ and 150~eV if NSA models are used), and one in which the
star surface is at uniform temperature (of $T_\infty \sim$80~eV, NSA) 
with an absorption edge at $\sim$0.65~keV present in the spectrum 
(for a discussion about the possible discrimination based on other 
arguments, see \S~\ref{disc_1}). 

We notice that, in the BB+BB and NSA+NSA models, the contributions of the 
two thermal components cross near $\sim$0.65~keV, which is probably the 
reason why the spectral fit can be accomodated by introducing a feature 
around this energy. This is at odds with \cite{sh09}, who proposed as a 
best-fitting model a NSA plus emission line: including a Gaussian emission 
line in the NSA model \cite[with an energy fixed at 0.53~keV, the value 
derived by][]{sh09} improves the fit only slightly (see 
Table~\ref{tabspec}) and the line equivalent width (EW) of 12~eV is 
smaller than that derived from the \CXO\ spectrum (28 eV).
Instead, we found that the fit improves by including an additional feature
in absorption (the energy resolution is not sufficient to differentiate between
an absorption edge or a negative Gaussian profile).
Furthermore, we also found that both models based on  two thermal 
components seem to further improve by introducing an absorption feature
(either, edge or Gaussian line) at $\sim 0.65$~keV. 
For instance, by adding  an absorption edge to the BB+BB model (with fixed
$N_H$) gives an F-test statistic value of 7.94,  and 
chance probability of $4\times
10^{-4}$.\footnote{
We warn that, although widely used, the F-test is not strictly
speaking a suitable method to test the significativity of (narrow)
spectral lines
and edges. This numbers must therefore be treated as an indication only.  
Given the uncertainties in modelling the continuum, further assessment of 
the significance of the feature (e.g. through Monte Carlo simulations) is 
not warranted.}

Finally, we tested a 2BB+PL model. The reason for this is that
there are some data points in the pn spectrum above 3 keV,
although they have
low significance (the background dominates at these energies, and
they could also be caused by improper background subtraction). The 
PL photon index was fixed at 2. We found that the PL contribution 
to the
total flux is $\sim8\%$. Again, the fit results in an increased $N_H$ as
compared to the 2BB case. We conclude that the existence of a
PL non-thermal component up to a flux level of about 10\% can not be
excluded. In order to have further insights on this issue, spectra with
better signal to noise ratio (S/N) above 3~keV are needed.

\subsection{XMM-Newton timing analysis}
\label{tim}

Given the rather different observing modes and sampling time
among the three instruments ($\Delta t$ of $\sim$5.67\,ms, 1.75\,ms and
2.6\,s for pn, MOS2 and MOS1, respectively) we first used all the event
lists
extracted from the three instruments during the two observations
together, and assumed the largest (MOS1) sampling time in order to look
for signals with periods larger than 5.2\,s. Moreover, in
order to maximise the sensitivity of the coherent pulsation search we 
minimise to two
(one for each observation) the number of averaged power spectra. Significant
power spectrum peaks were searched for by using the algorithm described in
\cite{IsraelStella96}.
No significant peak (above a 3.5$\sigma$ confidence threshold on
2.097.152 frequency trials)
has been found. The corresponding 3$\sigma$ upper limit on the pulsed
fraction (defined as the
semi--amplitude of modulation divided by the mean source count rate) is
about 10\% in the 5.2-10$^3$\,s range \cite[consistent, although 
tighter, 
with the constraints set by][]{sh09}.

\begin{figure}
\hspace{0.1cm}
\vbox{ \hbox{
\psfig{figure=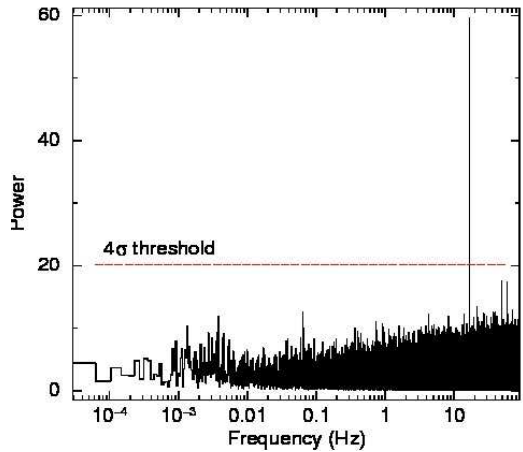,height=6cm,angle=-90}}
\caption
{EPIC pn and MOS2 power spectrum together with the 4$\sigma$ detection
threshold (dashed line). The power peak corresponding to the
$\sim$59\,ms signal is evident.
\label{power}}}
\end{figure}

The signal search was then carried out by using the pn and MOS2 event
lists only and with a
time resolution of 5.67\,ms in order to sample periods as short as
$\sim$12\,ms.
A highly significant peak ($\sim$\,11.5\,$\sigma$; see Fig.\,\ref{power})
was found at a frequency of 16.89242(2)\,Hz, corresponding to a period
of 59.19816(7)\,ms (uncertainties refer to the intrinsic Fourier
resolution of the power spectrum in Figure\,\ref{power}). Such a short and 
strictly
coherent period could be only accounted for by the spin period of 
a rotating NS.

\begin{figure}
\hspace{0.1cm}
\vbox{ \hbox{
\psfig{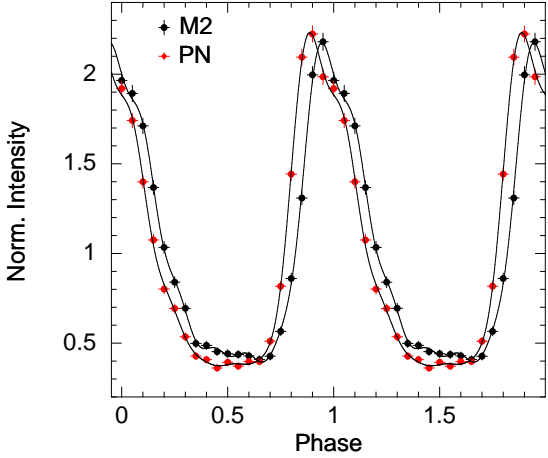}}
\caption
{EPIC pn and MOS2 (M2) background-subtracted light curves of PSR B1509-58
folded
at the best inferred period of $\sim$150\,ms. The solid lines represents the best fit obtained by adopting a model 
with a sinusoidal plus four harmonics.
\label{comp}}}
\end{figure}

A more accurate determination of the period, for each XMM 
observation, can 
be obtained by fitting the phases of the modulation over different time 
intervals. However, we noticed that the
simultaneous MOS2 and pn light curves folded to the above period show a
phase shift of about 0.10 and 0.15 (for the first and second
observation, respectively), making any results inferred from
the  phase fitting analysis unreliable. 
The presence of both phase shift and pulse distortion between MOS and pn
was also detected in other relatively fast pulsars, such as PSR B1706-44
\cite[with a pulse period of 102\,ms;][]{mcgow04}. In order to further
check for the presence of the observed phase shift we also analysed the
data from another pulsar, namely PSR B1509-58 
(pulse period of $\sim$150\,ms),  observed by \XMM\ in September 2000 with 
the pn
and MOS2 in the same  observational modes also set 
for our observations. The pn and MOS2 folded light curves of PSR B1509-58 
are reported in
Figure\,\ref{comp}.
The phase shift is statistically significant and equal to 0.050$\pm$0.001, 
corresponding to 7.5$\pm$0.2\,ms. Therefore, in the following we 
decide to use 
only the
pn data given that no calibration to assess the absolute timing accuracy 
of the
MOS exists (see the latest \XMM\ EPIC technical
reports\footnote{http://xmm2.esac.esa.int/docs/documents/CAL-TN-0018.pdf
and http://www2.le.ac.uk/departments/physics/research/src/ 
Missions/xmm-newton/technical/mallorca-2008-04/ 
08\_04\_09\_mallorca\_mos\_timode\_mgfk.pdf}), and because of the 
pn larger statistics.

In order to obtain a refined value of the period  we divided the first pn 
observation (obs. ID: 0601180101) in four time intervals of
duration  $\sim$\,5000\,s and inferred the phase of the modulation in
each interval \cite[see][for more details]{simone03}. The
scatter of the phase residuals was consistent with a strictly periodic
modulation at the best period of $P$=59.19821(1)\,ms or
$\nu$=16.892404(5)\,Hz (90\% confidence, 55094 MJD).
Unfortunately the inferred accuracy is not enough to keep the phase
coherency between the two observations which are separated by about 40
days (the inferred period uncertainty of the first \XMM\ observation 
translates
to a time uncertainty, at the epoch of the second pointing, of
$\sim$220\,ms corresponding to almost four period cycles). Correspondingly, we  inferred a
period separately for the pn dataset of the second observation (obs. 
ID: 0601180201) resulting
in a best value of $P$=59.19822(1)\,ms or $\nu$=16.892399(4)\,Hz (90\%
confidence), consistent with the period of the observation 0601180101.
The corresponding 3$\sigma$ upper limit on the first period derivative
$\dot{P}$ is $<9\times10^{-15}$\,s\,s$^{-1}$. 
\begin{figure}
\hspace{0.1cm}
\vbox{ \hbox{
\psfig{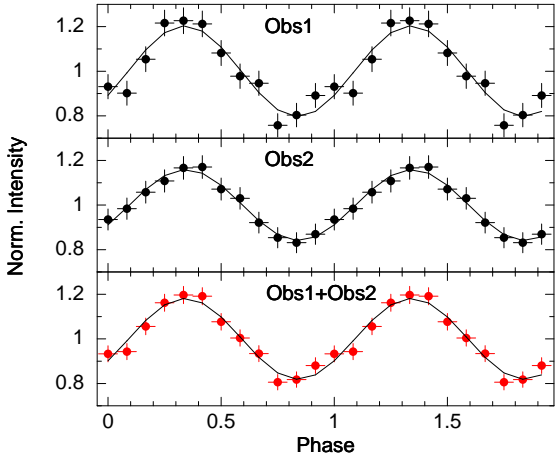}}
\caption
{EPIC pn background-subtracted light 
curves folded at the best periods
(see
text for details) for the first \XMM\ observation (upper panel, 
obs. ID: 0601180101), the
second observation (middle panel, obs. ID: 0601180201) and for the 
whole dataset together (lower panel).
\label{efoldX}}}
\end{figure}

No significant variations have been detected 
between the two observations (see Figure\,\ref{efoldX}). 
The signal shape is almost sinusoidal with pulsed
fractions of 18$\pm$3\% (90\% confidence, 
defined as $(C_{max}-C_{min})/(C_{max}+C_{min})$, where 
$C_{max}$, $C_{min}$ are the maximum and minimum number of counts). The 
upper limit on the presence of a second harmonic is of the order of 5-7\% 
(3$\sigma$ confidence level).

Finally, we searched for pulsation in the \CXO/HRC-I
archive data (obs. ID: 8508) with the aim  of obtaining
another period measurement and therefore a measure of the $\dot P$ or a
more constraining upper limit on it. 
No coherent signal was found, and the corresponding 3$\sigma$ upper limit 
on the pulsed fraction is larger than 100\%.

\subsection{Gamma-ray data analysis and results}
\label{datafermi}

Gamma-ray observations could represent an important step
to unveil the nature of Calvera.
In particular, the detection of pulsed emission at gamma-ray
energies can in principle be useful to constrain the period derivative
(and then the  rotational energy loss rate)
thanks to the large data spans typically provided by the current
generation
of gamma-ray telescopes with time tagging
accuracy as low as a few $\mu$s.

Hence, we searched for possible gamma-ray counterparts of Calvera
in the catalogue of gamma-ray sources \citep{abdoetal10a}
detected by the LAT pair-production
detector on board the \Fermi\ Gamma-ray Space telescope 
\citep{atwoodetal09}, and in the AGILE gamma-ray
bright source catalogue \citep{pittorietal09}.
No entries (spatial detections) in the gamma-ray catalogues
are compatible with the X-ray source position within
3$\sigma$ errors.

Nevertheless, it was worth performing a timing analysis on the LAT data, 
since the search for source timing signatures can be more sensitive than 
the spatial analysis to detect and evaluate the flux of weak periodic 
sources \cite[see e.g. the case of PSR B1509-58][]{abdoetal10b}, though
typically giving higher statistical errors on the flux. In order to 
perform gamma-ray 
timing analysis we retrieved
the available LAT public data on Calvera (photon data and
spacecraft data) through the \Fermi\ Science Support Center
(FSSC) web data-server 
interface\footnote{http://fermi.gsfc.nasa.gov/ssc/.}.
We analysed LAT data collected from 2008 August 4 (beginning
of science phase) to 2010 April 27 (LAT runs 239557417
through 294038606, where the numbers refer to the Mission
Elapsed Time (MET) in seconds since 00:00 UTC in 1 January
2001). During most of this time \Fermi\ was operated in sky
scanning survey mode (viewing direction rocking 35$^\circ$
north and south of the zenith on alternate orbits).

We performed LAT standard data processing using \textsc{gtselect},
\textsc{gtmktime} and \textsc{gtbary} tools obtained from the HEADAS
distribution of the \Fermi\ ScienceTools (version \textsc{v9r15p2}) built 
on 
Scientific Linux 5 64 bit operating system.
We selected E$>$100 MeV events extracted within the ``region
of interest'' (ROI) of 2 degrees from X-ray source position
\cite[S/N ratio of \Fermi-LAT pulsars is typically maximised
taking ROI of $\sim$1-2 degrees around pulsar position;][] 
{abdoetal10cat}.
We selected only events with high probability of being photons
(``diffuse'' event class=3, data quality=1) collected within
105 degrees maximum zenith angle (applying ROI-based zenith angle
cut) and excluding  data acquired during passages through the
South Atlantic Anomaly.
Solar System ephemeris JPL DE405 were used for the barycentric
corrections.
With the above selections, we obtained 2518 counts for the whole
$\sim$1.8 years data span.

\begin{figure}
\vbox{ \hbox{\psfig{figure=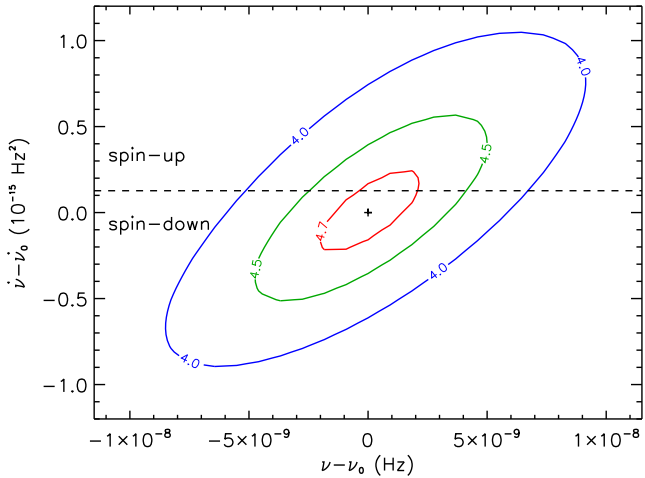,width=1.1\hsize}}
\caption
{The $\nu - \dot \nu$ contour plot for the
the gamma-ray timing
solutions with Z-test significance $>$4 sigma.
The most significant detection
is
$\nu_0=16.892401975(2)$~Hz, $\dot \nu_0 =-1.2(7) \times
10^{-16}$~Hz$^2$ (timing epoch 55094 MJD). The horizontal dashed line 
corresponds to $\dot 
\nu=0$. 
See text for details.
\label{gamma1}}}
\end{figure}

The search for the Calvera timing signature in gamma-ray started
from the ephemeris provided by the
X-ray observations ($\nu$ and upper limit on $| \dot \nu |$).
Standard epoch folding was performed within 10$\sigma$ errors from
the X-ray ephemeris.
A grid of gamma-ray frequencies and frequency derivatives was explored with
steps  oversampling by a factor of 10 the canonical resolution allowed by
the data span, $\nu_{res}$=$1/T_{span}=2 \times 10^{-8}$~Hz and $\dot 
\nu_{res}$=$2/T_{span}^2 =7
\times 10^{-16}$~Hz$^2$, where $T_{span}$ is the
time span of the data. Pearson's $\chi^2$ statistics was applied to the
10-bin folded pulse profile resulting from each period search trial.
Bin-independent parameter-free $Z^2_n$-test statistics
\citep{buccherietal83} was also applied to the data.

A significant pulsed signal from Calvera was detected both by 
$\chi^2$-test
and $Z^2_n$-test ($\nu=16.892401975(2)$~Hz, $\dot \nu =-1.2(7) \times
10^{-16}$~Hz$^2$,
$P=0.059198212396(7)$~s, $\dot P =4(2) \times
10^{-19}$~\ss, $Z^2_n$-test=26.2,
$\sigma$=4.7 with n=1 harmonics).
A reference epoch of 55094 MJD is assumed for the gamma-ray timing
analysis.
Weighting the corresponding detection probabilities with the
number of independent $\nu$ and $\dot \nu$ trials
($n_{trials} \sim$100) the overall
gamma-ray pulse significance is  $\sim 3.7\sigma$, corresponding to
a fake detection probability of only 3$\times$10$^{-4}$. We verified that 
our analysis procedure does not produce fake detections
even considering $\nu$ and $\dot \nu$ ranges much larger
($n_{trials} >10^4$)
than those compatible with the X-rays ephemeris.
In obtaining our timing solutions, the position of the source was held at
the  \CXO\  coordinates reported in \S~\ref{pos}. We have checked
that the positional uncertainty ($\leq$1 arcsec) does not significantly affect
barycentric corrections and then the rotational
parameters resulting from our timing analysis.

Fig.~\ref{gamma1} shows the $\nu - \dot \nu$ contour plot related to
the gamma-ray timing
solutions with Z-test signficance $>$4 sigma. Due to  $\nu - \dot \nu$
timing
solution degeneracy, as it is evident from the plot, positive frequency
derivatives cannot be in principle excluded although unlikely 
(we also note that the deep optical limit allows to rule 
out accretion from a binary companion) and only
upper limits on $ | \dot \nu | < 10^{-15}$~Hz$^2$
($ | \dot P | < 5 \times 10^{-18}$~\ss) can be
conservatively claimed. 
In case of spin-down, this corresponds to a 
rotational energy loss
$\dot E_{rot} <$10$^{33}$~\ergs. If
the spin-down is caused by magneto-dipolar braking the upper limit on  the
magnetic field is
$B<$5$\times$10$^{10}$~G.

Obviously, tight constraints on the ephemeris will rely on
longer data spans. In particular, $>$4$\sigma$ signal-to-noise
ratio on the $\dot \nu$ measurements is expected from 1 year of further
\Fermi-LAT observations.

\begin{figure}
\vbox{ \hbox{
\psfig{figure=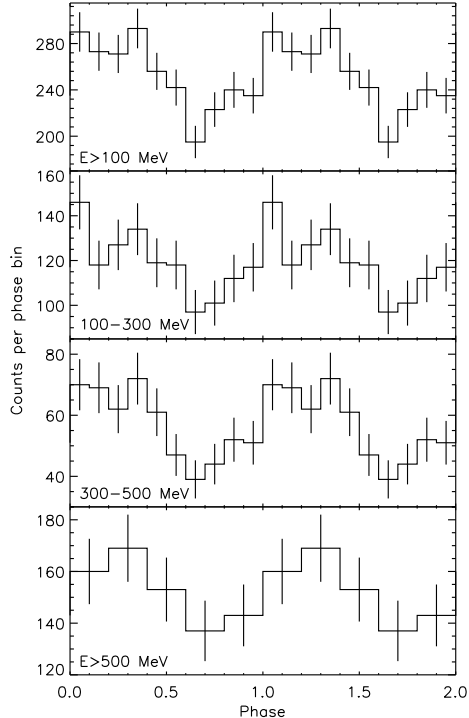,width=0.8\hsize}}
\caption
{Energy resolved gamma-ray lightcurves (total counts) taken with
\Fermi-LAT.
The pulsed signal is detected at $\sim$5$\sigma$ in the whole E$>$100 MeV
band ($\sim$6~ms bin resolution).
Gamma-ray emission from Calvera seems present at least up to $\sim$1 GeV
though the pulsed signal significance is low at E$>$500 MeV.
\label{gamma2}}}
\end{figure}

The  energy-resolved gamma-ray light curves related to the most 
significant
solution (corresponding to the cross in Fig.~\ref{gamma1}) are shown in
Fig.~\ref{gamma2}. A single broad peak is detected in the gamma-ray light-curves
without significant pulse shape variations as a function of energy.

The pulsed flux was computed considering all the counts above the minimum
of the light curve, using the expression $PF=(C_{tot}-n
N_{min})/Exp \equiv C_{pul}/Exp$, where
$C_{tot}$ is the  total number of counts, $n$ is the number of bins in the
light curve, $N_{min}$
are the counts of bin corresponding to the minimum, and $Exp$ is the
exposure in cm$^2$s units.
This method is ``bin dependent'', but we have checked that different
(reasonable) choices of both the number of bins (i.e., $n >10 $) and 
the
location of the bin centre (10 trial values were explored) do not significantly affect the results.
The pulsed counts for $E >$100~MeV are $C_{pul}=455 \pm 95$~cts
($C_{pul}=275 \pm 75$ for
$E >$300~MeV) corresponding to  $\sim$15\% of total gamma-ray counts
(the unpulsed flux can be ascribed to diffuse gamma-ray background
emission). Comparable results are obtained using the the expression
adopted for the calculation of the X-ray pulsed fraction 
$(C_{max}-C_{min})/(C_{max}+C_{min})$.
No pulsed signal can be significantly ($>$3$\sigma$)
disentangled when data are restricted to
narrower ranges in the low energy part (e.g. 100-300~MeV), although a
modulation (in
phase with the profiles obtained from broader energy bands) is still
visible in the light curves (Fig.~\ref{gamma2}). Therefore, only
rough gamma-ray
spectral measurements can be performed with the present count statistics.

We made a LAT exposure cube from the spacecraft data file using 
\textsc{gtltcube} 
procedure from \Fermi\ ScienceTools and created the exposure map 
using the \textsc{gtexpcube} tool. The resulting total exposure of the 
whole data 
span at the
source position for $E>100$~MeV is $2.2 \times 10^{10}$~cm$^2$s
($5.4 \times 10^{10}$~cm$^2$~s
for $E >300$~MeV),  corresponding to a $E >100$~MeV pulsed flux
$PF=(4.1 \pm 0.9) \times 10^{-8}$~ph/cm$^2$/s ($PF=(6.4\pm
1.7)\times 10^{-9}$~ph/cm$^2$/s for $E >300$~MeV).
We corrected the pulsed flux accounting for the additional source counts
falling outside the ROI (2 degrees radius) estimated according to the
instrument
Point Spread Function (PSF). We also verified that the pulsed fraction  
systematic errors due to
spectral uncertainties in the exposure calculation are below count statistics
errors.
A comparison of $E> 100$~MeV and $E>300$~MeV pulsed fluxes provides a 
rough gamma-ray
photon index estimate $\alpha =2.5 \pm 0.5$ (100~MeV-10~GeV).

In the frame of a spin-powered pulsar interpretation, the resulting
$E> 100$~MeV luminosity,  assuming a beaming angle of 1~steradian,
is $L_{\gamma}=1.3 \times 10^{32} d_{kpc}^2$~\ergs,
where $d_{kpc}$ is the source distance in kpc. The observed 
gamma-ray luminosity
and
the upper limit on the spin-down power ($<$10$^{33}$~\ergs) constrain the
distance below $\sim 1$~kpc, considering a likely gamma-ray conversion
efficiency upper limit of 10\%\footnote{An efficiency of $\sim 10\%$ is indeed very high, 
although it may be attained in some sources. As an example, \cite{mig10} 
recently revised the distance of PSR~1055-52, finding a relatively low 
value 
of $\sim 300$~pc. This gives a gamma-ray efficiency as high as 10\%, in 
line with the value assumed in this paper.}.

\begin{table*}
\caption{Pulse phase spectroscopy}
\begin{tabular}{lcccccccc}
\hline
\multicolumn{3}{c}{ } &
\multicolumn{3}{c}{Low-temperature BB component} &
\multicolumn{3}{c}{High-temperature BB component} \\
Model & $\chi^2$ / d.o.f.& $N_H$      & $kT_1$    & F$_{bol}^{max}$ & F$_{bol}^{min}$ & $kT_2$     & F$_{bol}^{max}$ & F$_{bol}^{min}$ \\
      &                  &  10$^{20}$ &       & 10$^{-13}$      & 10$^{-13}$      &            & 10$^{-13}$      & 10$^{-13}$      \\
      &                  &  cm$^{-2}$ & eV    & \ergscm2             & 
\ergscm2         & eV     & \ergscm2         & \ergscm2         \\
\hline
PPS1  & 1.17/242 &  4.2$\pm1.0$       & 96$\pm10$ & 8.15        & 6.32        & 246$\pm13$ & 7.37        & 5.71        \\
PPS2  & 1.01/241 &  4.3$\pm0.7$       & 96$\pm10$ & 7.54        & 7.42        & 246$\pm12$ & 8.04        & 5.06            \\
\hline
\end{tabular}

Bolometric
fluxes for the BB components derived from the EPIC pn spectrum of 
observation 0601180101.
\label{tabpps}
\end{table*}

\subsection{XMM-Newton Pulse phase spectroscopy}
\label{pps}

\begin{figure}
\vbox{ \hbox{\psfig{figure=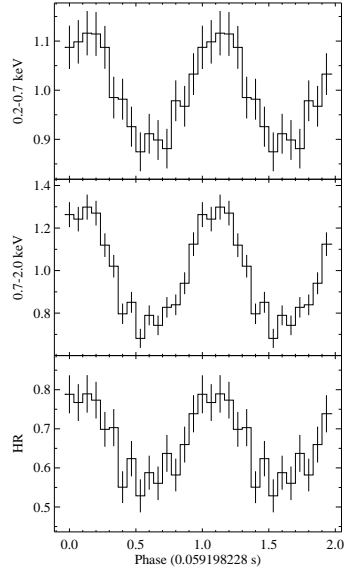,width=0.6\hsize}}
\caption
{Background subtracted folded light curves obtained from the combined EPIC 
pn data of both observations in two different energy bands. The pulse 
profiles are normalised by the corresponding average count rates of 0.197 
cts~s$^{-1}$ and 0.133 cts~s$^{-1}$ (0.2-0.7  keV and 0.7-2.0 keV, 
respectively). The bottom panel shows the hardness ratio $HR = 
CR(0.7-2.0~{\rm keV})/CR(0.2-0.7~{\rm keV})$, where $CR$ denotes the count 
rate in 
the corresponding energy band.
\label{hrfig}}}
\end{figure}

In order to look for spectral variations in the X-ray flux of Calvera 
we first created background subtracted light curves in the energy bands 
0.2-0.7 keV and 0.7-2.0 keV and a hardness ratio by dividing the count 
rates in the  hard band by those in the soft band. We used single- and 
double-pixel events (PATTERN=0-4) from the EPIC pn data, combining both 
observations and extracted events from the same source and background 
regions as used for the extraction of the spectra. The light curves were 
folded with a period of 59.198228~ms, derived from a simple chi-square 
folding test of the data, assuming a constant period. The resulting pulse 
profiles and the hardness ratio as function of pulse period are shown in 
Fig.\ref{hrfig}. 

\begin{figure}
\psfig{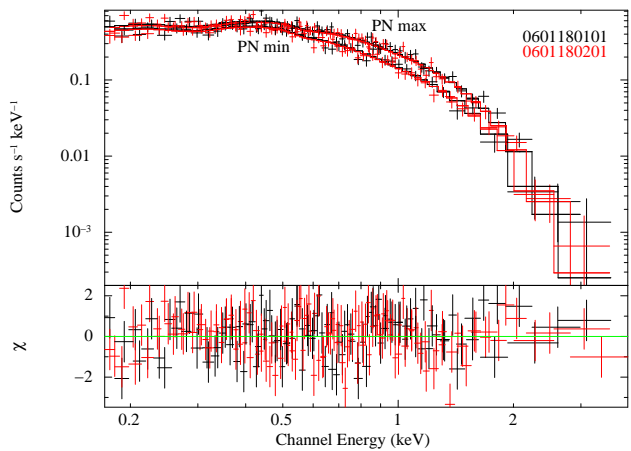}
\caption
{EPIC pn spectra from two different pulse phase intervals covering pulse maximum and minimum
from the two \XMM\ observations. The histograms show the double blackbody model PPS2 as
decsribed in the text and Table~\ref{tabpps}.
\label{figpps}}
\end{figure}

We then investigated the possibility of a changing relative contribution
of the two spectral components with spin phase. To this aim, we have 
produced EPIC pn spectra (MOS1 has insufficient time resolution for pulse 
phase spectroscopy, PPS)
around pulse maximum and minimum (each covering 0.5 in phase) and fitted
the four spectra (two observations, two phases) with the double BB model.
Similarly to the case of the pulse average spectra, we fitted all the pn 
spectra
simultaneously. We performed fits with two model flavours: models
PPS1 and PPS2.
In model PPS1 only the relative overall normalisations are allowed to vary 
between 
spectra. This
corresponds to a case without variation in spectral shape between the
two phase intervals.
In PPS2 we allow different normalisations for all the BB components, i.e.
a variation of the relative contribution of the two blackbody components is possible.
The derived spectral parameters and bolometric fluxes are summarised in
Table~\ref{tabpps}, while spectra are shown in Fig.~\ref{figpps}.
Again, because of no significant long-term flux variations, we only 
report
in the table the fluxes inferred from the pn spectra from 
the first observation (obs. ID: 0601180101).

Both  models assume no temperature changes with pulse phase, and provide a fully
acceptable fit. Given the current statistics, we are not in the 
position to allow
more free parameters in the fit or to test different spectral models.
As one can see from Table~\ref{tabpps}, model PPS2 yields a 
better
description of the spectral behaviour with pulse phase (F-test statistic 
value of 41.3 and chance probability $6.9 \times
10^{-10}$). This shows that the spectral shape changes with pulse 
phase, as also indicated by the variations in the hardness ratio 
(Fig.\ref{hrfig}), and suggests that
the relative contribution of the hotter component is higher during pulse maximum
and lower during minimum.

\section{Radio observations}
\label{radio}

Following the discovery of pulsations in the X-rays and gamma-rays, the
X-ray position of Calvera was searched for radio pulsations using the
Effelsberg telescope with the aim of better understanding the nature of 
the
source and in particular, 
in case of detection,  to measure the period derivative given the 
higher precision generally achievable in the radio band. 
Observations were taken on
MJD 55330 at 1.36 GHz with a bandwidth of 240 MHz spread over 410
frequency
channels using a sample time of 55 $\mu$s for 78 minutes and for four 
epochs (a total of 312 minutes). On
MJD
55337 a further observation was taken using the same setup for 60 minutes.

These observations were folded using the X-ray ephemeris and searched over
a
range of frequency dependent time delays to correct for the unknown amount
of dispersion (caused by free electrons along the line-of-sight) between
Calvera and Earth. These dispersion measures ranged from 0 to 1000 
cm$^{-3}$~pc and the step size was optimised for a pulsar with a period of 
$\sim 
60$~ms. The observations were also searched blindly for any periodicities and
for single dispersed pulses (though impulsive radio frequency
interferences at Effelsberg make this latter type of search very
difficult).

No radio pulses from Calvera were detected down to a sensitivity of 
$S=0.05$~mJy
for each of the observations assuming a duty cycle of 5\%. These
observations improved the previous flux density upper limit
\cite[obtained by][with the Westerbork Synthesis Radio Telescope]{he07}
of
almost an order of magnitude \cite[by assuming a spectral  
index of 1.7, see][]{kra98}. Assuming a distance $d < 1$ 
kpc, this
translates into an upper limit of 0.05 mJy kpc$^2$ for the
pseudo-luminosity
(defined as $S \times d^2$) at 20cm, a limit that encompasses 99\% of all
known pulsars (and 100\% of the mildly recycled ones to which Calvera
could
belong, see \S~\ref{disc_3}) reported in the version 1.40 of
the ATNF Pulsar Catalog\footnote{\rm
{http://www.atnf.csiro.au/research/pulsar/psrcat/}}. 

Note that the Effelsberg observations are only sensitive to pulsations 
and
not to continuum emission; inspection of archival NRAO VLA Sky Survey data
\citep{con98} showed no clear evidence of point-like or extended
radio sources in the vicinity of Calvera (see \S~\ref{diff}).

\section{Search for diffuse emission in the field of Calvera}
\label{diff}

\begin{figure}
\vbox{ \hbox{
\psfig{figure=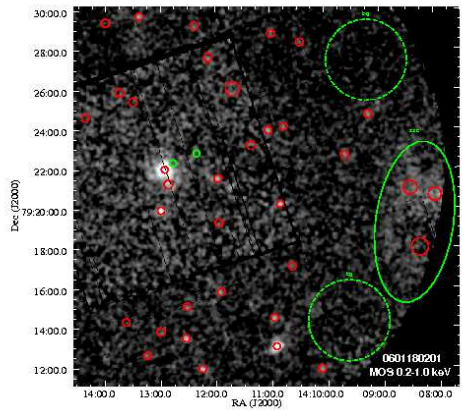,width=0.95\hsize}}
\caption
{MOS1+MOS2 image in the 0.2-1.0 keV band. The
spectrum of the extended source  west of Calvera has been extracted from  
the
region inside the green ellipse. The two
green dashed circles
indicate the regions used to extract the background spectrum.
Small green and red circles are as in Fig.~\ref{figfield},
right panel. \label{figext}}}
\end{figure}

\begin{figure}
\psfig{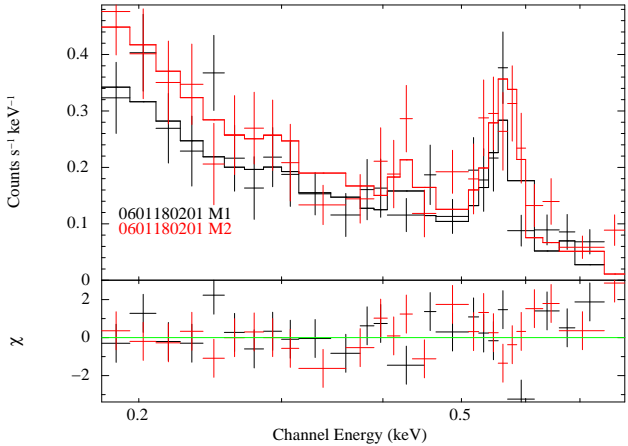}
\caption
{Spectral fit of MOS1/MOS2 data from the extended emission with a NEI
model (see text for details).
\label{specext}}
\end{figure}

In order to search for possible sources of diffuse emission in the
proximity of Calvera, we analysed in detail the \XMM\ and \ROSAT\ all-sky 
survey (RASS) images.

Using \XMM\ MOS1 and MOS2 data, we created colour images for  the second 
\XMM\ observation. At low energies, different features are 
clearly visible.
North of Calvera we can see an area with increased soft emission, which
however shows a sharp edge at the border of the CCD, and therefore it is
likely to be related to higher noise in one of the MOS CCDs.
More interestingly, the image shows an extended region of soft X-ray emission
west of Calvera (at the outer rim of the field of view, where the source
detection software also identifies three sources, see the combined MOS
0.2-1.0 keV image shown in Fig.~\ref{figext}).
The extended emission is relatively soft and best visible below 1.0 keV.

The extended emission is also clearly seen in the standard
RASS images, where it is best visible in the 0.1-0.4~keV band, 
but
also seen in the 0.5-2.0 keV band. The \ROSAT\ source detection also 
located
a source within the extended emission region: this detection is included
in the bright source catalogue \citep{vog99}, but lacks an optical
identification \citep{zic03}.
The \ROSAT\ images do not show significant emission north of Calvera,
confirming that the feature seen in the EPIC image is likely to be due to
CCD noise.

Although it is unlikely that the extended emission is related to Calvera,
considering their angular separation of 13\arcmin, it remains an 
interesting
detection. Its X-ray morphology is reminiscent of
an old supernova remnant, or a pulsar wind nebula.
Further insights on its nature can be obtained by a spectral analysis.
Therefore, we extracted the MOS1+MOS2 spectra from an
elliptical region (246\arcsec and 117\arcsec semi-axes) covering the 
extended
source. The background was extracted using two circular regions 
(122\arcsec
radius), free  of \XMM\ detections (see Fig.~\ref{figext}).
Spectra were fitted using a non equilibrium ionisation (NEI) model
(Fig.~\ref{specext}), which reproduces the X-ray spectra of faint
SNRs, e.g. in the Magellanic Clouds \cite[see
e.g.][]{fil08}.
We found that the spectrum of the diffuse emission is well
represented by this model ($\chi^2_r = 1.35 $ for 46 d.o.f.), and
clearly shows the (unresolved) emission line triplets of O VII and N VI.
From the fit,  O appears to be over-abundant by a factor of $\sim 2-3$.
Unfortunately, due to the scarce data, the temperature is
unconstrained (it has been  fixed at 0.7~keV during the fit) and we
could only derive an upper limit on the column density,
$N_H< 1.5 \times 10^{20}$cm$^{-2}$ ($3 \sigma$). 

\begin{figure*}
\vbox{ \hbox{
\psfig{figure=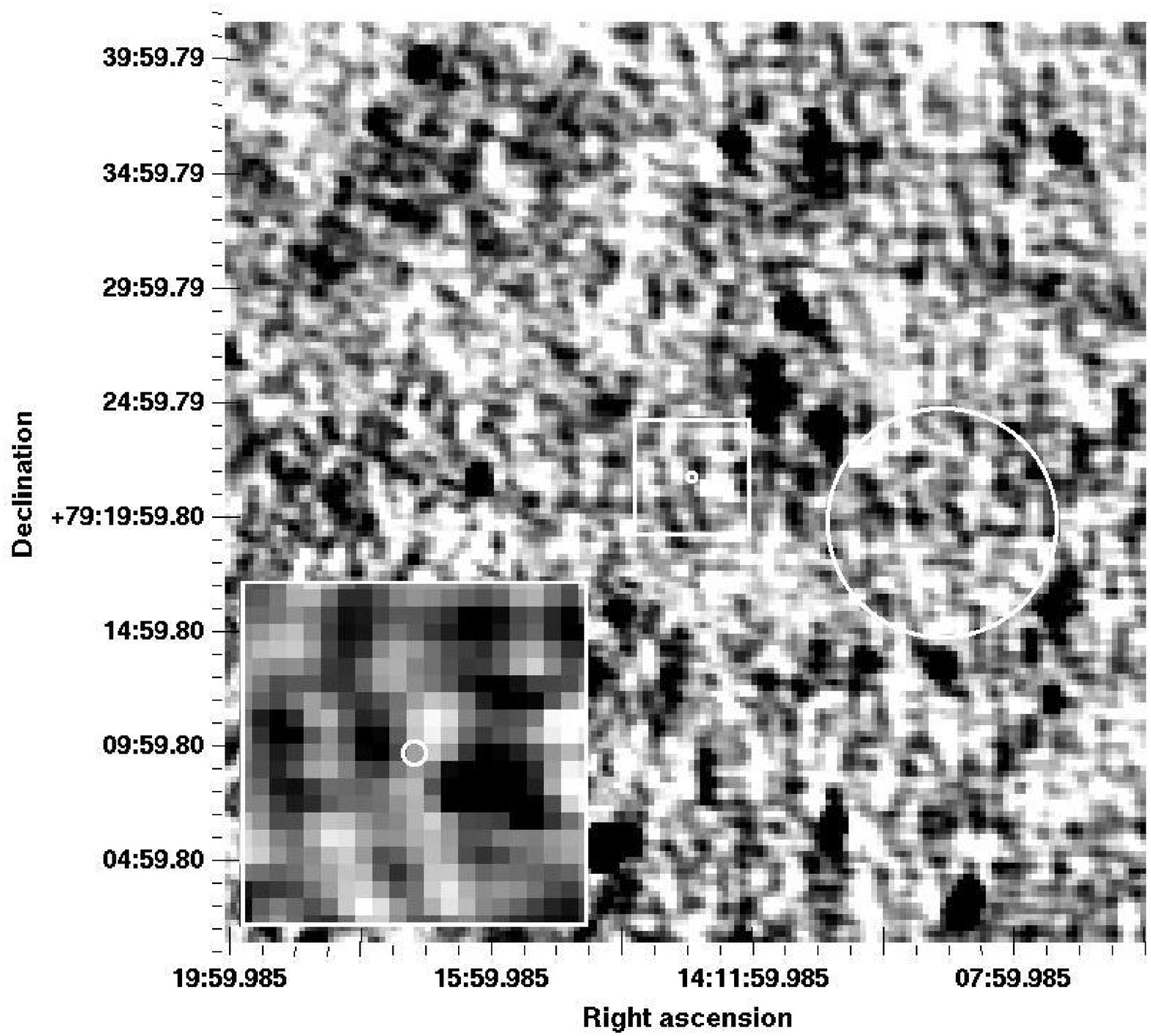,width=0.53\hsize}
\hspace{0.02\hsize}
\psfig{figure=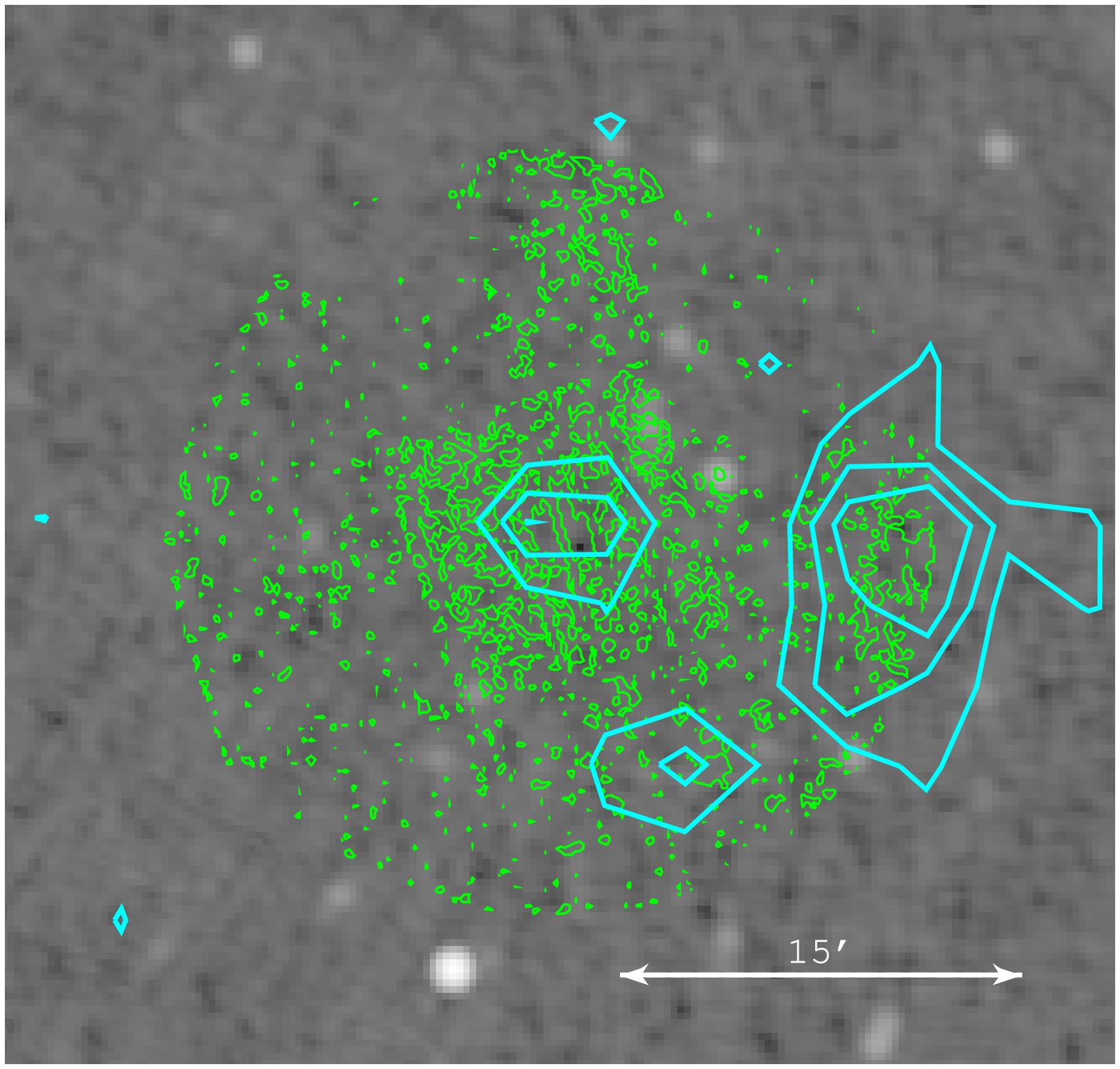,width=0.47\hsize,angle=0}}
\caption
{{\it Left panel:}  NVSS image of the Calvera field.  The contrast has 
been increased to show faint sources. The position of
Calvera is marked by  the  white circle whose radius has been fixed to
$10\arcsec$ for a better visualisation.  The closest radio source is
detected $\sim 30\arcsec$  northwest of the Calvera position and is well
away from the  \CXO\  error circle. The  white square ($5' \times 
5'$)
marks the region highlighted in the inset (lower left corner). The large
white circle ($5 \arcmin$ radius) is drawn around the position of the
extended source seen in the \XMM\  and  \ROSAT\ data. No extended
radio emission  is detected on angular  scales larger than $\sim 
2\arcmin5$
down to a flux limit of $\approx 1.5 \mu$J arcsec$^{-2}$.
{\it Right panel:} Same as in the left panel, with the  X-ray
contours from \XMM\  (green) and RASS (cyan) overplotted. For \XMM\ we 
used the 
combined
MOS 0.2-1.0 keV
image (just one contour level is shown,  corresponding to $5.4  
\times 10^{-3}$~cts s$^{-1}$ 
arcmin$^{-2}$) and for RASS the total 0.1-2.4 keV
image (three contour levels: (2.8, 4.1 and $5.5)\times 10^{-3}$~cts 
s$^{-1}$ 
arcmin$^{-2}$).
\label{nvss}
}}
\end{figure*}

In order to
investigate further the nature of the extended emission, we inspected the
DSS R, B and IR images, but did not find evidence for spatial structures.
We  also checked  for evidence of  diffuse $H_{\alpha}$ emission at the
position
of the \XMM\  extended source using  images from the The
Virginia
Tech Spectral-Line Survey (VTSS)\footnote{http://www.phys.vt.edu/~halpha/},
which has pixel size of  1\arcmin6 and field of  view of $10^\circ \times
10^\circ$ per pointing. Unfortunately, the Calvera field is not yet
available in the processed Survey data. No other $H_{\alpha}$  survey data
are available, which cover the field with a sufficient spatial resolution.

We then inspected the field in radio continuum using data from the 
NRAO/VLA Sky
Survey (NVSS) at a frequency of 1.4 GHz, which covers the sky  at
$\delta > -40^{\circ}$ with beams of $45\arcsec$ FWHM \citep{con98}.
From
the
NVSS data (Fig.~\ref{nvss}) we could not find evidence for any radio
source, either
point-like or extended, at the position of Calvera, down to a flux level
of $\sim 0.1$~mJy ($3 \sigma$) at 1.4~GHz. As far as the radio emission
from Calvera itself is concerned, this upper limit is much less
constraining than those discussed in \S~\ref{radio}.
The closest radio source detected
in the NVSS data is located at $\sim 30\arcsec$ from the position of
Calvera, it is not extended and has an estimated flux of $\sim 0.47$ mJy.
At the same time, we found about a dozen radio sources ($\sim 
0.5-2.5$~mJy)
within a circle of $5\arcmin$ radius centred on the position of the 
X-ray extended emission. However, none of them shows evidence of
extended radio emission on angular scales larger than $\sim 2\arcmin5$. We
could not find evidence of diffuse emission over the whole searched area
down to a limit of $\approx 1.5$ $\mu$J arcsec$^{-2}$, which we assume as
the upper limit on the radio brightness of the  extended emission detected
in the \XMM\  and \ROSAT\ data.

\section{Discussion}
\label{disc}

We have presented a new multi-wavelength study of
the NS candidate Calvera (1RXS J141256.0+792204), based on
two new \XMM\ observations, on one set of publicly available 
\Fermi-LAT 
data
and on one new observation taken at Effelsberg in the radio band. 
We also used \CXO\ and \ROSAT\ archival data. 
We
discussed previous determinations of the source position, and concluded
that at present the best
estimate is that based on absolute  \CXO\  ACIS-S astrometry, i.e.
$\alpha=14^{\rm h} 12^{\rm m}
55\fs84$,
$\delta=+79^\circ 22\arcmin 03\farcs7$ with a  90\% confidence
error of  0\farcs6.

The combination of \XMM\ and \Fermi-LAT has allowed us to discover the 
period of the source, $P\simeq 59.19$~ms, and to place an upper limit on 
its spin-down rate, $ | \dot P | < 5 \times 10^{-18}$~\ss. We warn that 
positive frequency derivatives can not be excluded, although, based on the 
contours presented in Fig.~\ref{gamma1}, they appear to be less likely. 
This possibility will be addressed by future timing monitoring, and will 
not be discussed further hereafter. Instead, in case of spin-down our  
measurements correspond to a rotational energy loss $\dot E_{rot} < 
10^{33}$~\ergs, a large characteristic age $>1.55$~Gyrs, and 
a low magnetic field, $B < 5 \times 10^{10}$~G (under the 
assumption of magneto-dipolar breaking). 

Although the best-fit of the X-ray spectrum is not unique (see 
Table~\ref{tabspec}), according to our findings (see \S~\ref{disc_1} for a 
detailed discussion), the most likely spectral model consists of two 
thermal components (blackbodies) with temperatures $kT_1 \sim 150$~eV, 
$kT_2 \sim 250$~eV, plus possibly an absorption edge at $\sim 0.65$~eV 
($\sim 0.65 (1+z)$~eV when measured at the star surface).  
Similar spectral features are found in other classes of thermally emitting 
NS, as the XDINSs \cite[see e.g.][and references therein]{tur09} or 
1E~1207.4-5209 \cite[e.g.][]{big03, mori06}, which, however, are 
characterised by a much larger spin period. By interpreting the edge as an 
electron or a proton cyclotron feature gives $B = 6 \times 10^{10} 
(1+z)$~G or 
$B= 6 \times 10^{13}(1+z)$~G, respectively: 
the former value is not too different 
from the limit inferred from the timing (which, however, only 
constrains the large scale dipolar component). Instead, we do not find 
substantial evidence for the emission line at $\sim 0.5-0.6$~keV proposed 
by \cite{sh09}; on the 
other hand, considering the thermal character of the underlying continuum, 
a feature in absorption appears a more likely outcome. 
The X-ray 
luminosity is $L_X \sim  10^{32} d_{kpc}^2
$~\ergs (with small variations depending on the assumed spectral model, 
see Table~\ref{tabspec}), and there is  no evidence for a PL tail up to 
a flux contribution of $\sim 10\%$. 

\begin{figure*}
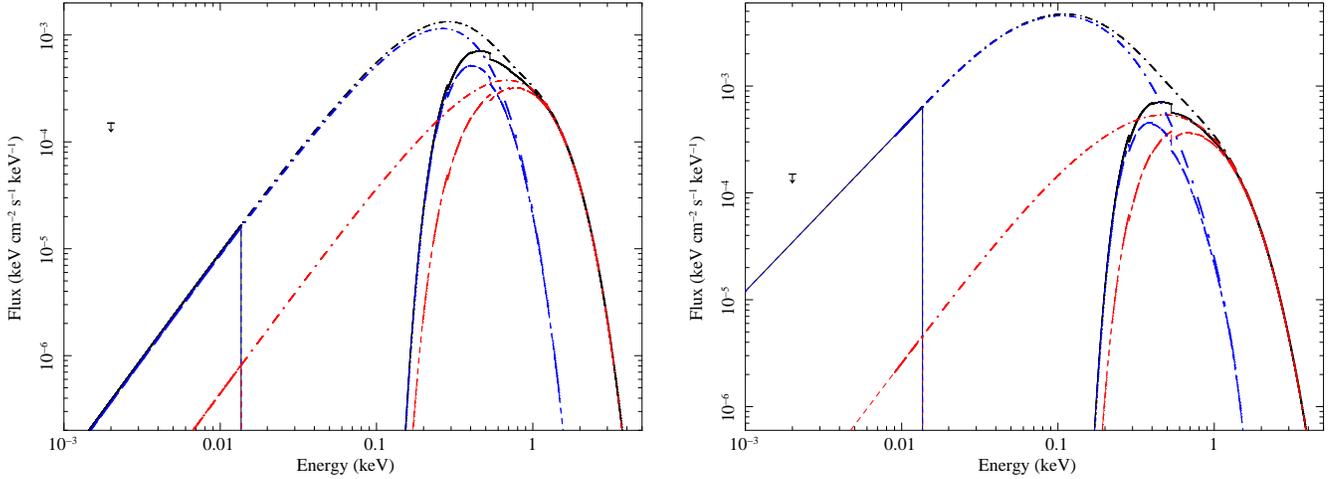

\vbox{ \hbox{
\psfig{figure=fig15.ps,width=0.48\hsize,angle=270}
\hspace{0.02\hsize}
\psfig{figure=fig16.ps,width=0.48\hsize,angle=270}}
\caption
{{\it Left panel:} The best-fitting double
blackbody model  shown in Fig.~\ref{fig1} (left panel) extrapolated in the
optical band. Red lines: hot component, blue lines: cold component, black
lines: total. Solid and dashed-dotted lines correspond to
absorbed and unabsorbed spectra. The arrow marks the $g$-band upper limit inferred
from Gemini-North+GMOS imaging (see text and Rutledge et al. 2008).
{\it Right panel:} Same as in the left panel for the NSA+NSA model shown
in the right panel of Fig.~\ref{fig1}.
\label{fig2}}}
\end{figure*}

Overall, Calvera is a very puzzling pulsar. It has a thermal soft
X-ray spectrum (with a possible absorption feature),
but, at variance with other similar sources as the XDINSs, it is not a
``soft'' source: through our timing search we found that it is still
detected at $>4 \sigma$ even at $E>300$~MeV. A rough estimate of the
100~MeV-10~GeV photon index gives $\alpha=2.5 \pm 0.5$.  
The gamma-ray luminosity is $L_{\gamma}=1.3 \times 10^{32} 
d_{kpc}^2$~\ergs, 
computed by assuming a 
beaming angle of $1$~steradian. Assuming a gamma-ray conversion efficiency 
upper limit 
$<$10\%, and assuming that the 
observed gamma-ray luminosity is powered by spin-down, the upper limit 
on the rotational energy loss ($< 10^{33}$~\ergs) provides a quite strong 
constraint on the distance, that turns out to be below $\sim 1$~kpc (much 
lower distances might result from the future estimate of the true value of 
$\dot E_{rot}$).

The source is not detected in the optical and radio bands. 
Based on the non detection of the source on a Gemini-North+GMOS imaging of
the field, \cite{rut08} reported a $3\sigma$ upper limit of $g > 
26.3$~mag,
or $F_g < 0.11$~$\mu$Jy, at  4750~A. As for the radio band, our  
Effelsberg observation allowed us to set a deep upper limit on the source 
pulsed emission of 0.05~mJy, assuming a duty cycle of 5\%.

Finally, we report evidence for diffuse X-ray emission in the field 
of Calvera, $\sim 13\arcmin$ west of the source. 

In  the following, we will first discuss the spectral and timing 
results, and then we will elaborate on the implications of our new findings   
regarding the perspectives on the source interpretation.

\subsection{The new timing and multi-wavelength spectral properties}
\label{disc_1}

Thanks to the high sensitivity of \XMM\ at low energies, we have been
able to reach a good characterisation of the X-ray spectrum.
In agreement with previous findings
based on  \CXO\  data \citep{sh09}, we found that the X-ray spectrum
can
not be reproduced by a single component model, while a satisfactory fit can
be obtained by adding a spectral feature to an atmospheric model (NSA).
This spectral decomposition has the advantage that it requires a
relatively low value of interstellar absorption ($N_H \approx 2.6 \times
10^{20}$~cm$^2$), compatible with that inferred from H\textsc{I} maps in 
the source
direction \citep{kal05, sh09}.
In this case, the star surface emits at a uniform temperature of $\approx
100$~eV, and the feature is more likely to be an absorption edge at
$\approx 0.65 (1+z)$~keV.  

The main problem with a scenario based on uniform surface emission is that
our measurement of a relatively large X-ray pulsed flux 
($PF \sim 18\%$)  seems to
rule out a relatively smooth thermal map. Pulse profiles produced by the 
thermal surface distribution induced by a
simple core-centered dipolar magnetic field have been investigated long   
ago by \cite{page95}, under the assumption that the surface emits
(isotropic) blackbody radiation. Because of gravitational effects and of
the smooth temperature distribution (the temperature monotonically
decreases from the poles to the equator), the pulse modulation is quite
modest ($PF \la 10\%$). 

Indeed, our spectral fits show that the spectral decomposition
is not unique: an equally likely possibility is
that the spectrum is made of two
thermal components (again with possibly a spectral feature at $\sim
0.6-0.7$~keV), reflecting the presence
of two zones of the NS surface at different temperatures. The two
components can be modelled either with a NSA or a BB, with
the difference that, since
atmospheric models are harder than a blackbody at the same temperature,
the temperatures inferred from a NSA+NSA fit are systematically lower
than those obtained from a double BB model. As a consequence, a BB+BB and
a NSA+NSA decomposition result in very different predictions for the
amount of flux expected in the optical band (see for example
Fig.~\ref{fig2}), and  in principle it may be
possible to use optical observations to discriminate between the two 
options. Unfortunately, in practice, this is unfeasible.
The $g$ band upper limit from the  Gemini-North+GMOS imaging translates 
into $\approx 1.5
\times 10^{-4}$~keV cm$^{-2}$s$^{-1}$keV$^{-1}$ at $\approx 2 \times
10^{-3}$~keV, which
is still compatible even with the large fluxes expected in the atmospheric
emission scenario (see Fig.~\ref{fig2}, right panel). In order to use this
approach to
robustly discriminate among X-ray spectral models, we would need an upper
limit on the
optical flux at least 1-2 orders of magnitude lower, which requires deep
imaging of the field down to $g\sim 29-31$. This is clearly unfeasible 
with
the current instrumentations.
\begin{table} \setlength{\tabcolsep}{0.02in} \centering
\begin{tabular}{lcccc} \hline Model & $N_H$ & $R_{cold}/R_{hot}$ &
Distance & Comment\\
 &   & km &  pc & \\
\hline
BB + BB &  free &   3.1 / 0.43  & & $N_H>N_{gal}$ \\
BB + BB &  fixed &    1.9 / 0.36  & &  \\
(BB + BB)$\times$edge &  fixed &    1.6 / 0.36  & &  \\
NSA + NSA &  free &     & 175 & $N_H>N_{gal}$ \\
NSA + NSA &  fixed &     & 1550 &  \\
(NSA + NSA)$\times$edge &  fixed &     & 2250 &  \\
\hline
\end{tabular}
\caption{Radii of the emitting regions and source distance inferred from
the
spectral modelling with two thermal components. In the case of the BB+BB
fit, the reported values for the radius of the hot and cold component
correspond to a  source distance of 1~kpc; for smaller distances they
should be scaled down accordingly. In the case of the NSA+NSA models, the
source distance is computed by assuming that the colder component is
emitted by the whole star surface, and
using $R_{star}=12$~km and
$M_{star} = 1.4 M_{sun} $. We warn that, in this case, the inferred 
distances depend on the assumed values, although slightly. For instance, 
would the mass be as large as  $2 M_{sun} $, the distance will turn out 
to be a factor 
$\sim 15-17\%$ smaller. The interstellar absorption is left free to
vary
during the fit or fixed at the Galactic value; values are reported in
Table~\ref{tabspec}.}
\label{tab2} \end{table}

In this respect, some insights may be obtained exploiting the detection of 
the source at high energy based on the LAT timing analysis. In fact, the 
fits to X-ray data with two thermal components using a NSA+NSA or a 
BB+BB model result in different estimates of the distance and of the size 
of the emitting regions (see Table~\ref{tab2}). 
In Table~\ref{tab2}, we report the predicted distances inferred by the 
fits  based on a double NSA model, for the limit case of the colder 
component emitted from the 
whole NS surface. This is an overestimate for the area of the colder 
region, since a second, hotter NSA component is present. However, the 
ratio of the cold/hot areas is $\approx 0.02-0.05$ (similar to that 
implied by the BB+BB fit), 
and our estimates of the distance would not change significantly by 
assuming that $\sim 95\%$ (instead of 100\%) of the whole surface is 
emitting at the lowest temperature.
In this case, if $N_H$ is fixed at a value
compatible with the Galactic absorption, the distance turns out to be
quite large,  $\sim$1.5-2~kpc, which seems too large when compared
with the limit inferred from the gamma-rays (unless the gamma-ray
conversion efficiency
is extremely high).  Similarly, the BB+BB model
is only
compatible with emission from the whole surface (from the coldest 
component) if the star is at 5~kpc or
more, which, again, contradicts the gamma-ray limit.

In fact, a BB+BB model is compatible with the relatively large PF only for 
a $d\sim 1$~kpc, which gives quite small BB radii 
($\sim 0.5-3$~km) suggesting a scenario in which the two
thermal components
originate in two different small spots at the star surface. 
If this is the case, the rather small emitting areas, together with 
a  pulsed fraction of $\sim 18\%$, and the sinusoidal pulse 
shape, are suggestive of a geometry in which the star is a quite aligned 
rotator seen at a large inclination angle. This is also supported by the 
pulse-phase spectroscopy and hardness ratio variation, which 
indicate that there is only a moderate 
spectral evolution with phase. This may be explained if both 
emitting caps are always (partially) in view as the star rotates, with the 
hotter spot becoming more visible near the pulse maximum. 
The gamma-ray pulses are also quite 
sinusoidal, with no evidence for changes at different energies and a 
pulsed fraction similar to that in the X-rays. However, the current 
impossibility of phase-align the X- and gamma-ray observations 
(see \S~\ref{tim} for all details) prevents to 
reach any conclusions on the relative positions of pulse maxima.

\subsection{The nature of the source}
\label{nature}

Our new findings, in particular the new discovery of the X-ray and 
gamma-ray period and the measurement of an upper limit for the spin-down 
rate, unequivocally identify Calvera as a
relatively fast
spinning NS, and
provide long-sought crucial information to shed light onto the
conundrum about its nature (see below
and \S~\ref{storia}). 

Based on the preliminary spectral informations available at that time,
\cite{rut08} attempted to classify Calvera by comparing it with different
classes of compact objects that occupy different regions in the 
luminosity/effective temperature/emitting radius
spaces. In light of the new discovery of a relatively 
fast period, some of the previously proposed scenarios as that of
an XDINS or a  magnetar are now ruled out. Similarly, Calvera's spin 
period is too large to be compatible with an interpretation in terms of 
a (fully) recycled millisecond pulsar. 

\begin{figure}
\hspace{0.1cm}
\vbox{ \hbox{
\psfig{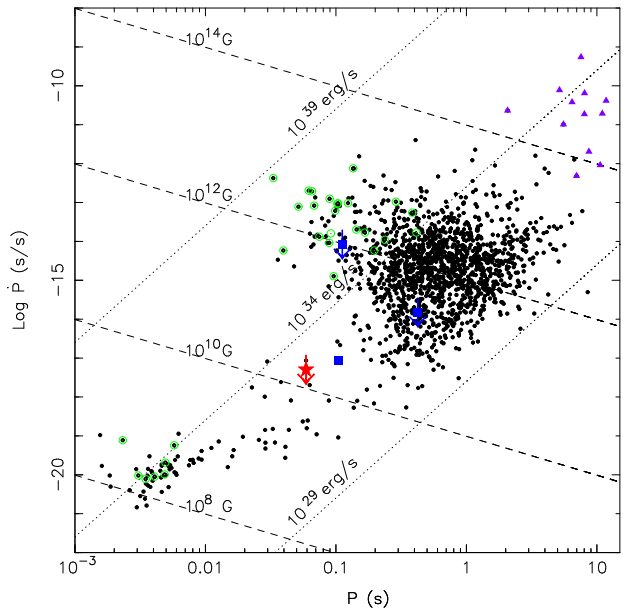}}
\caption{$P-\dot P$ diagram for Galactic field NSs: the red 
  star marks our upper limit for Calvera, black dots are radio pulsars 
(from the ATNF pulsar
  catalogue; Manchester et al.~2005), the blue squares are CCOs with a
  measured (or an upper limit value of) $\dot P $ (Gotthelf \&
  Halpern, 2007, 2009; Halpern \& Gotthelf 2010) and violet triangles
  are magnetars (data from Mereghetti, 2008). The NSs encircled in green  
are sources 
detected by \Fermi\ (Abdo et al. 2010a). Dashed lines indicate equal 
magnetic 
field, while dotted
  lines indicate equal value for the rotational energy losses, $\dot 
E_{rot}$.}
\label{martafig}}
\end{figure}

In the context of the $P$-$\dot P$ diagram, Calvera 
appears as one of the transition objects that populate the zone between
the bulk of the pulsar crowd and the group of recycled
millisecond pulsars (in the bottom-left corner of
Fig.~\ref{martafig}). 
Also, the rotational parameters (hence the values
of $\dot E$ and $B$) are consistent with those of the CCOs, the most
similar of which is the source in Kes 79.  

These findings, taken together with the fact that the source is apparently 
isolated and it is not associated with a SNR (but see \S\ref{disc_2}), 
open exciting novel  
possibilities, among which the most likely are that of an ``orphan CCO'' 
or that of a 
mildly recycled source that was once member of a high mass binary system 
(see \S~\ref{disc_2},~\ref{disc_3}).

\subsubsection{A new CCO?}
\label{disc_2}

Based on the timing properties and on the  thermal character of the 
spectrum, a first possibility is that Calvera is  a new CCO. In this 
respect, we 
note 
that the X-ray energetics of the source is also unusual. The upper limit 
on $\dot E_{rot}$ ($<10^{33}$~\ergs) is only a 
factor of 10 larger than the 
X-ray
luminosity inferred from \XMM\ data ($\sim 10^{32}$~\ergs at 1~kpc). 
While an efficiency of $\sim 10\%$ in
gamma-ray is not unusual (see \S\ref{datafermi}), the conversion
efficiency in the X-ray band is typically lower, of order of $\sim   
10^{-2}-10^{-3}$ \citep{beck10}. If the X-ray emission if 
powered by the source spin-down (for instance in the case in 
which it originates from two hot spots at the star surface, heated by 
back-flowing magnetospheric currents),  
a similar efficiency would imply a very low distance, of order of $\sim 
100-300$~pc (or even less, 
considering that our estimate of the upper limit on $\dot E_{rot}$ is 
conservative). 
Alternatively, the thermal X-ray emission may have a different origin 
(e.g. anisotropic surface cooling). In this case we note that, based on 
our spectral 
fitting, we cannot exclude 
that a $\sim 10\%$ of the X-ray flux is ascribed to a non thermal 
PL component; if this fraction of the total X-ray emission is 
spin-down powered, an 
efficiency of $10^{-2}$ translates in $d \sim 1000$~pc. 

Interestingly, apart from binary systems and magnetar sources, that are 
thought to be
powered, respectively, by accretion or by the super-strong magnetic
field, the only sources that may be characterised by $\dot
E_{rot}\la L_X$ are the (slowly rotating) XDINSs 
\citep{kvk09} and the CCOs. For instance, 
the only CCO for which the spin period derivative has been (recently)
measured \cite[i.e.  PSR J1852+0040 in Kesteven 79, see][]{ha10}, has
$L_X \sim 10 \dot E_{rot}$. In all other cases only an upper limit on
the period derivative is available; still, 1E~1207.4-5209 has $L_X \sim
2 \times 10^{33}$~\ergs and $\dot E_{rot}< 10^{32}$~\ergs
\citep{go07}. Puppis A has $\dot E_{rot}< 10^{33}$~\ergs, that does
not exclude an efficiency of $\sim 10^{-2}$ \citep{goha09}.
The X-ray emission of CCOs is thermal and characterised by a double
black body spectrum, without the presence of hard X-ray tail, similar
to the X-ray spectrum Calvera. 

A possible counter-argument for the CCO interpretation might be raised by 
comparing with Fig.~2-3 by \cite{rut08}: Calvera's surface temperature 
is quite low
with respect to those typically observed in this class, and, 
since the typical
X-ray luminosity of other CCOs is $\sim 10^{33}$~\ergs, this
interpretation would require a source distance of 2-3 kpc (against the 
constraint
from the gamma-rays). Nevertheless,  we warn   
that these problems with a CCO interpretation may be not too severe. 
The comparison of Calvera with CCOs based on the
temperature comes with caveats, since it does not account for the 
fact 
that  different
CCOs might have different ages, or that the  
thermal history of a CCO
might have been modified by a recent phase of accretion. 
Also, 
 we 
note that the derived CCO luminosities are based on the estimated
distances of the host SNRs, which are usually affected by
uncertainties of the order of 50\% or more. Thus, using CCO
luminosities as standard candles could be risky.   

The possible association with a  SNR, or lack thereof,  is a
crucial piece of information to determine whether Calvera is, or not, a
CCO. We already 
noticed that the extended
X-ray source detected west of Calvera in the \XMM\ and \ROSAT\ data
shows evidence for triplets of O VII and N VI in the spectrum, which
are robust signatures of SNRs (although a final confirmation could   
only come from H$\alpha$/S II/O III ratios). If we assume that Calvera
has moved 13' away from the centre of this putative SNR, the kinematic age 
of Calvera (and in turn the age of the SNR) results $t = 39000 
d_{kpc}/v_{100}$~yrs, where 
$v_{100}$ is the source velocity in units of 
100 km\,s$^{-1}$. This is much lower than 
the 1.55~Gyr age estimated from the spin
parameters. However, this is a common
characteristic of CCOs: the lower limits on their spin-down ages
(although smaller than that of Calvera) are always much
larger than the SNR ages. A way out, within the ``anti-magnetar''
interpretation \citep{ha10}, is to assume that the initial
spin periods of these pulsars were very close to their current values, so
that they have not changed much during the pulsar's lifetime. In
other words, the
spin-down age is not representative of the true age of the source but only
reflects its initial conditions at birth.

Despite these considerations, we consider 
it unlikely that this extended emission is associated to
Calvera. 
At a distance of $\sim 1$~kpc a typical CCO-associated SNR is
expected to be much larger: for instance the 7~kyr old SNR associated 
to 1E~1207.4-5209 would appear with an angular size of $\sim
2^\circ/d_{kpc}$. Instead, the size of the extended emission derived  
from the  \ROSAT\ contours is $16\arcmin \times 
8.5\arcmin$, which suggests that the SNR is a background object. 
We note that the upper 
limit on the column density inferred from the SNR spectrum ($N_H< 1.5 
\times 
10^{20}$cm$^{-2}$)  is smaller than the Galactic $N_H$ in 
the direction of Calvera. This would make the extended emission 
likely closer, however this comparison is hampered by the limited 
photon statistics of the diffuse source.  
Considering the angular separation and  
the low $\dot E_{rot}$ of the pulsar, it is also unlikely that the    
extended source is a pulsar wind nebula associated to Calvera.    

On the other hand, if Calvera is an underluminous CCO at a low
distance of only $\sim 300$ pc, the large spatial scale would make it
hard to detect the SNR with narrow field X-ray instruments like EPIC. 
Furthermore, it can  
not be ruled out that Calvera is an old ($\ge 1$~Myr) CCO and that its
host SNR has already expanded and faded away in the interstellar medium. 

\begin{figure}
\hspace{0.1cm}
\vbox{ \hbox{
\psfig{figure=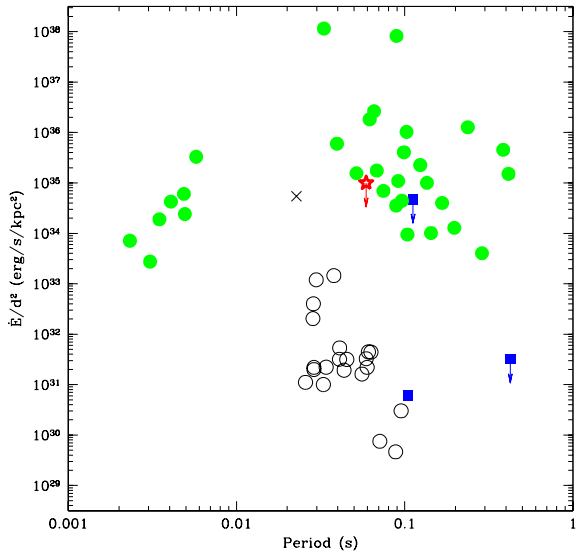,height=8cm}}
\caption{
Ratio $\dot E_{rot}/d^2$ plotted versus spin period for
various samples of NSs. The green filled dots represent all
the pulsars detected so far in gamma-rays 
by \Fermi-LAT 
(according to Abdo et al. 2010a). The black empty dots represent all
the radio pulsars in the ATNF pulsar catalogue having spin period
between 20 and 100 ms and surface magnetic field smaller than
$10^{11}$ Gauss (i.e. those generally interpreted to be mildly
recycled pulsars; none of them has been detected so far by \Fermi-LAT
in gamma-rays). The red star represents the position of Calvera, 
based on our upper limit on  $\dot E_{rot}$ and on a (low) distance of 
$100$~pc. The blue filled
squares refer to the three CCOs with a known value (or an upper limit)
for the spin period derivative and hence for $\dot E_{rot}$. The black
cross reports the position of the mildly recycled pulsar in the Double
Pulsar binary PSR~J0737-3039, for which there is a tentative detection (to 
be
confirmed) of pulsed emission in gamma-ray with {\it Agile-GRID} 
(Pellizzoni, in preparation), and no published detection with \Fermi-LAT. 
All
the distances in this plot (but for the three CCOs) have been derived from
the best-estimate reported in the ATNF pulsar catalogue at 30 June
2010 (Manchester et al. 2005). In most cases, they are inferred from
the dispersion measure of the radio pulsar.}
\label{martafig2}}
\end{figure}

This ``orphan CCO'' scenario is potentially intriguing, especially in
view of the possible connections between old CCOs and other NS
classes.  Still, in this scenario Calvera would be a quite extreme
case.  So far, no CCO has been detected at gamma-ray energies. This   
may be due to observational limits: firstly, in cases in which the  
embedding SNR dominates it is difficult to resolve the gamma-ray
emission from the CCO.  Secondly, detecting CCOs as weak gamma-ray
pulsars is difficult because only three of these sources have an X-ray
period and only one (Kes 79) has a measured $\dot P$. 
In Fig.~\ref{martafig2} we report 
the ratio
$\dot E_{rot}/d^2$ (a often used figure of merit for the detectability in
the gamma-ray band of a source with $\dot E_{rot}<10^{34}$ erg s$^{-1},$
e.g. \citealt{abdoetal10cat}) plotted against the spin period for
various relevant samples of neutron stars (see caption of
Fig.~\ref{martafig2}). 
As we can see, among the CCOs with a measured (or an upper 
limit value of) $\dot P $, two of them have a relatively low value of 
$\dot E_{rot}/d^2$, which may hamper the gamma-ray detection.  
Would a CCO 
nature be confirmed, Calvera will represent a unique case and our
timing technique, here applied to \Fermi\ data, constitute a key tool to
investigate the high energy spectrum of other members of the class (the 
most promising candidate for this search being Puppis A, whose upper 
limit for $\dot E_{rot}/d^2$ is, as in the case of Calvera, compatible 
with relatively a large value).

\subsubsection{Is Calvera the first discovered mildly recycled
gamma-ray pulsar?}
\label{disc_3}

The spin period and the upper limit on the
spin period derivative place Calvera well within the
area of the $P-\dot P$ diagram which is known to be populated by the
so called mildly recycled radio pulsars.  Hence the plausible and
very intriguing possibility that Calvera represents the first case
ever of a mildly recycled gamma-ray pulsar.

It is commonly believed that recycled radio pulsars are formed when a
relatively old neutron star is spun up by accretion while in a binary
system \citep{bis74}. The amount of mass accreted, hence of spin up,
is related to the duration of the X-ray binary phase, and ultimately
depends on the mass of the companion: low mass X-ray binaries are
thought to be the progenitors of the so called ``fully'' recycled
radio pulsars (often referred to as millisecond radio
pulsars\footnote{By comparing again with Fig.~2-3 in the paper by
\cite{rut08}, we can see that in principle, if the source distance
were low ($\sim$few 100s of pcs), the properties of Calvera inferred
from the BB+BB fit (i.e. typical temperatures around 100-200~eV and
rescaled blackbody radii of $\sim 0.1-0.2$~km) might be compatible
with those of the fully recycled radio pulsars observed in 47
Tucanae. However, the now measured spin period rules out the
millisecond radio pulsar scenario proposed by \cite{rut08}.}) with
spin periods of few ms, while binary systems with an intermediate or
an high mass companion give rise to mildly recycled radio pulsars with
periods of tens of ms (e.g. \citealt{lor04}).

In this picture we note that the lack of a detection of Calvera in the 
radio band (down to a very deep flux density limit) can be 
plausibly explained by the relatively narrow radio beam missing the 
Earth. As 
to the X-ray luminosity, one can apply similar considerations to those 
reported for the CCO hypothesis (see \S~\ref{disc_2}), i.e. a 
rotation-powered nature requires a distance as low as $\sim 100-300$~pc.
The double BB X-ray spectrum, however, is somehow unusual for a
recycled pulsar, since most of them are characterised by either
single component (either BB or PL) or  BB+PL spectra. 

The observability in the gamma-ray band is an unprecedented feature
for NSs with the rotational parameters of Calvera and one
may wonder if it is due to some peculiar emission process operating in
this source or to observational biases. In this respect we first note
that, provided the distance is around or below $300$~pc, the gamma-ray 
efficiency nicely fits with the average
value (around few percent) seen in the gamma-ray pulsars observed by
\Fermi-LAT. Even assuming a larger distance of 1~kpc, the
gamma-ray efficiency remains compatible with what observed in at least
one other object \citep{mig10}.  Also, the gamma-ray spectrum is
within the range of those shown by the \Fermi\ pulsars, although
pointing toward the tail of the steepest spectral indices in the
spectral parameter distribution. 
By inspecting Fig.~\ref{martafig2}, we can see that the large majority of 
the NSs with rotational parameters in the range of those of the
mildly recycled radio pulsars (empty dots in the diagram) have a
$\dot E_{rot}/d^2$ ratio well below the minimum value which resulted in a
detection with \Fermi\ so far (filled dots,
\citealt{abdoetal10cat}). Although the location of Calvera in this plot 
is highly unconstrained, we notice that in principle its 
$\dot E_{rot}/d^2$ ratio can be at least $\sim 10$ times larger (for a
distance less than 300 pc) than that of the best cases among the
mildly recycled radio pulsars   
(similar results hold when plotting the analogous parameter $\sqrt{\dot 
E_{rot}}/d^2$, e.g. \citealt{abdoetal10cat}). 
Therefore, in the framework of the
interpretation of Calvera as a mildly recycled gamma-ray pulsar (for
which a short distance is largely preferred), it seems that the
relatively large observed gamma-ray flux of Calvera is due to its
proximity rather than to a stronger intrinsic emission with respect to the
other known rotation powered neutron stars.

It is worth noting that the relatively steep gamma-ray
spectral index of Calvera could be echoed in the spectrum of the
mildly recycled pulsar in the Double Pulsar binary
(PSR~J0737$-$3039A, \citealt{bur03}; \citealt{lyne04}). For the latter
(whose $\dot{E}/d^2$ is also very promising, see Fig.~\ref{martafig2})
there is a tentative detection with {\it Agile-GRID} (Pellizzoni, in 
preparation) and no detection so far with \Fermi-LAT,
which would indicate a very soft gamma-ray spectrum (interestingly
enough, also the X-ray spectrum of the Double Pulsar appears unusually
soft among the recycled pulsars, \citealt{pellizz08},
\citealt{poss08}). If confirmed by future analysis, this may be a 
peculiarity of the high
energy emission from the class of mildly recycled pulsars.

As to the evolution of Calvera, the presence of a main sequence, or a
giant or a white dwarf companion can be ruled out by the deep optical 
upper limit. This in turn suggests that Calvera is not
the descendant of an intermediate-mass X-ray binary, since the end
products of these systems usually comprise a neutron star orbiting a
heavy Carbon-Oxygen white dwarf or \cite[in some peculiar cases, see
e.g.][]{li02} a Helium white dwarf. Given the considerations
above, it seems very likely that Calvera was recycled in a high-mass
X-ray binary, in which also the companion star eventually experienced
a supernova explosion.  The most probable outcome of this event is the
disruption of the binary, with the release of two isolated neutron
stars, one having the typical parameters of a ordinary pulsar and the
other - usually dubbed disrupted recycled pulsar, DRP
\cite[i.e. see][]{lor04} - having a moderate spin rate (in the range of
few tens of ms) and a surface magnetic field which is intermediate
between that of the ordinary pulsars and that of the fully recycled
pulsars. Alternatively, there is the possibility that the binary
system survives the supernova explosion, leading to the formation of a
double neutron star (DNS) binary.
If the absence of a binary companion will be supported
by future dedicated campaigns of observation\footnote{Our 
present 
data do not allow us a search for orbital periodicities.}, Calvera will enter the
rare class of the DRPs, which would make its discovery particularly
worthwhile for investigating various still open issues on the final
stage of the evolution of high-mass X-ray binaries and on the
supernova kick.

For instance, there is currently a large mismatch between the
theoretical expectations on the relative number of DRPs and DNSs and
the result of the observations. According to the fact that the
survival of the binary in the second supernova explosion requires
properly tuned parameters for the pre-supernova system and/or for the
kick associated to the supernova event, population synthesis studies
\cite[e.g.][]{portegies98} indicate that DRPs are expected to be
generated at a significantly higher rate than DNSs.  Since the
rotational parameters of the NSs are similar in the two
classes of objects, their lifetime in the radio band and their radio
emission properties should also be similar and hence one would expect
to detect up to $\sim$10 times more DRPs than DNSs (\citealt{lor04}).
At variance with this prediction, to date 8 DRPs in the Galactic field
are reported, a number matching that of the 8 known DNSs (from ATNF
pulsar catalog at 30 June 2010). Supernova kicks smaller than those
usually assumed in population synthesis have been proposed to at least
partially alleviate this problem: in fact that would reduce the gap
between the birth-rates of the DRPs and that of the DNSs.

Notably, the distances (derived from the dispersion measure) of the
known DRPs (all selected in the radio band) cluster mostly between 2
and 3 kpc, with the closest source (PSR~J2235+1506) located at 1.2
kpc. Therefore, Calvera  would 
turn out to be much closer than all the other DRPs, suggesting the
existence of a significant (if not even dominant) contribution of the
gamma-ray pulsars to the population of the mildly recycled neutron
stars.  Were this the case, the current observed ratio between the
number of radio selected DRPs and DNSs could not be representative of
the whole population. Of course, additional discoveries of mildly
recycled gamma-ray pulsars (with particular emphasis for blind
searches), as well as detailed binary pulsar population synthesis
models are necessary to assess the biases due to the small-number
statistics and to properly test the aforementioned hypothesis.  The
confirmation of a DRP interpretation for  Calvera and the discovery of 
similar sources would have strong implications: e.g. it
will impact on the estimates of the beaming factor (i.e. the fraction
of sky swept by the emission beam(s)) in the radio and gamma bands for
the mildly recycled neutron stars, thus constraining their emission
models; it will lead to increase the overall birth-rate of the
descendants from high-mass X-ray binaries in the Galaxy; it will also
suggest the existence of some DNS binary at close distance from the
Earth and which escaped radio detection so far. This may in turn lead
to an upward revision on the expected rate of events of merging of
DNSs in the Galaxy, a key prediction for the current generation of
ground-based gravitational wave detectors.

Once the spin down rate (and hence the surface magnetic field) of
Calvera will be measured it might be possible to estimate its initial
spin period \cite[post spin-up phase, see Fig.~2 in][]{lor04}, as well
as the time since the occurrence of the second supernova explosion in
the progenitor binary. With the present data we can only notice that,
in both the DRP and DNS hyphotheses, Calvera is expected to be much
younger than the age inferred from the spin-down rate
\citep{acw,lor04}: e.g. for our upper limit on $\dot{P}$, the time
elapsed from the second supernova would be $10^8$ yr, under the
hypothesis that Calvera was spun up by accretion of mass at the
Eddington rate up to the limit imposed by its magnetic field,
\citep{lor04}. Since the position of Calvera - slightly outside the
Galactic disk, at a distance $<0.6 d_{kpc}$ above the Galactic plane -
could be ascribed to the kick imparted to the neutron star (or to the
DNS binary) by the second supernova, a constraint on the time since
the explosion (even better if complemented with constraints on the
proper motion of the source) will in turn allow one to study the kick
imparted to the neutron star (or to the DNS binary) in the supernova.

\section{Conclusions} 
\label{conc}

Thanks to our multi-wavelength campaign, we have been able to recognise
in Calvera an intriguing, low magnetised pulsar with a relatively fast
period, $\sim 59$~ms. This rules out most of the previously proposed
scenarios for the nature of the source.
Calvera's properties are reminiscent of those of CCOs and mildly recycled 
pulsars, although other
interpretations can not be ruled out (mainly because our timing
solution is still compatible with a spin up scenario). On the other hand,
even in comparison with the other members of these two classes, the source
can be singled out thanks to its unique characteristics, mainly the
hard gamma-ray emission joined to a thermal X-ray spectrum.

The discovery of peculiar, possibly transitional, sources such as Calvera
is of the utmost importance to gain a full understanding of the Galactic
NS phenomenology and to eventually establish links between the different
classes into which isolated NSs have been catalogued up to now.  The
existence of bridging objects is progressively coming into view, as in the
case of PSR B1509-58 for the magnetar/high-field spin-powered PSRs
connection \citep{pel09}, of the newly discovered radio magnetar
\citep{levin10}, or the bursting young pulsar PSR~J1846-0258
\citep{gav08}. 
Furthermore, although Calvera is most likely to be
isolated, the current optical limits do not allow us to exclude a NS
companion, in which case the source would represent one of the rare DNS
systems discovered in the Galaxy.

Given the unique placement of Calvera in this framework, further
investigations aimed at better assessing its properties are definitely
warranted. Accurate X-ray timing can provide a positive determination
of the source spin-down rate, and a better characterisation of its
spectrum, especially concerning the presence of a high-energy PL tail
and of spectral features, is bound to reveal much on the processes
which power its X-ray emission, and ultimately on its true
nature. Presently we were able only to place an upper limit on the
pulsed radio flux. Calvera may be genuinely radio quiet but the search 
for radio emission must be pursued further since a
detection at radio wavelengths would yield an independent, accurate
determination of $\dot P$, the dispersion measure and the star proper
motion \cite[the latter may be also pursed in the X-ray band, 
see][]{sh09}. 
Follow-up X-ray and optical narrow-band imaging observations will allow one
to better characterise the properties of the putative SNR, and to 
constrain both its age and distance. All these 
issues are crucial to assess if there is indeed a
connection between Calvera and the nearby diffuse emission, the 
presence of which was first reported in this investigation. 
New observations at gamma-ray energies will allow one to better constrain the
source spin evolution, shed light on the spectral distribution at
gamma-ray energies and how the latter changes with rotational
phase. Finally, the search for gamma-ray emission from other sources with
similar properties (CCOs, mildly recycled and low $\dot E_{rot}$ PSRs,
...)  using the novel timing technique which led us to the detection
of Calvera in the gamma-rays is certainly in order and will be matter
of future work.

\section*{Acknowledgments}

The work of GLI and RT is partially supported by INAF/ASI through grant 
AAE-I/088/06/0. PE acknowledges financial support from the Autonomous 
Region of Sardinia 
through a research grant under the program PO Sardegna FSE 2007--2013, 
L.R. 7/2007 ``Promoting scientific research and innovation technology in 
Sardinia''. Based on observations obtained with \XMM, an ESA science 
mission with instruments and contributions directly funded by ESAMember 
States and NASA. We acknowledge the \Fermi\ Science Support Center at 
NASA/Goddard providing guest observers with smart public data retrival 
services. The 100-m Effelsberg radio telescope is operated by the 
Max-Planck-Institut fuer Radioastronomie of the Max-Planck-Society.
This research has made use of data obtained from the \CXO\
Data Archive and software provided by the \CXO\ X-ray Center
(CXC) in the application package CIAO. We are grateful to an anonymous 
referee for several helpful comments.

\label{lastpage}

\end{document}